\documentclass[11pt]{article}
\usepackage{amssymb,amsmath,amsfonts}
\usepackage{graphicx}
\usepackage{graphics}
\usepackage{eepic,epsfig}

\makeatletter \@addtoreset{equation}{section}

\makeatother

\begin{document}

\title{Induced fermionic currents in de Sitter spacetime \\
in the presence of a compactified cosmic string}
\author{A. Mohammadi$^{1}$\thanks{
E-mail: a.mohammadi@fisica.ufpb.br}, E. R. Bezerra de
Mello$^{1}$\thanks{
E-mail: emello@fisica.ufpb.br} and \thinspace\ A. A. Saharian$^{1,2}$\thanks{%
E-mail: saharian@ysu.am} \\
\textit{$^{1}$Departamento de F\'{\i}sica, Universidade Federal da Para\'{\i}%
ba}\\
\textit{58.059-970, Caixa Postal 5.008, Jo\~{a}o Pessoa, PB, Brazil}\vspace{%
0.3cm}\\
\textit{$^2$Department of Physics, Yerevan State University,}\\
\textit{1 Alex Manoogian Street, 0025 Yerevan, Armenia}}
\maketitle

\begin{abstract}
We investigate the vacuum fermionic currents in the geometry of a
compactified cosmic string on background of de Sitter spacetime. The
currents are induced by magnetic fluxes running along the cosmic
string and enclosed by the compact dimension. We show that the
vacuum charge and the radial component of the current density
vanish. By using the Abel-Plana summation formula, the azimuthal and
axial currents are explicitly decomposed into two parts: the first
one corresponds to the geometry of a straight cosmic string and the
second one is induced by the compactification of the string along
its axis. For the axial current the first part vanishes and the
corresponding topological part is an even periodic function of the
magnetic flux along the string axis and an odd periodic function of
the flux enclosed by the compact dimension with the periods equal to
the flux quantum. The azimuthal current density is an odd periodic
function of the flux along the string axis and an even periodic
function of the flux enclosed by the compact dimension with the same
period. Depending on the magnetic fluxes, the planar angle deficit
can either enhance or reduce the azimuthal and axial currents. The
influence of the background gravitational field on the vacuum
currents is crucial at distances from the string larger than the de
Sitter curvature radius. In particular, for the geometry of a
straight cosmic string and for a massive fermionic field, we show
that the decay of the azimuthal current density is damping
oscillatory with the amplitude inversely proportional to the fourth
power of the distance from the string. This behavior is in clear
contrast with the case of the string in Minkowski bulk where the
current density is exponentially suppressed at large distances.
\end{abstract}

\bigskip


\bigskip

\section{Introduction}

It is well known that the geometrical and topological effects play a central
role in a large number of physical problems. They have important
implications on all scales, from subnuclear to cosmological. In particular,
in quantum field theory the properties of the vacuum crucially depend on the
both geometry and topology of the background spacetime. In the present paper
we consider combined effects of the geometry and topology on the vacuum
current densities induced by magnetic flux tubes. As a background geometry
we consider de Sitter (dS) spacetime and the topological effects are induced
by two types of sources. The first one will correspond to a planar angle
deficit due to the presence of a cosmic string and the second one comes from
the compactification of the spatial dimension along the cosmic string.

The cosmic strings are among the most important types of topological defects
that may have been formed by the phase transitions in the early universe
\cite{Vile94}. Though the recent observations of the cosmic microwave
background radiation have ruled out them as the primary source for
primordial density perturbations, the cosmic strings give rise to a number
of interesting physical effects such as the doubling images of distant
objects or even gravitational lensing, the emission of gravitational waves
and the generation of high-energy cosmic rays (see, for instance, \cite%
{Damo00}). Recent developments on the formation of topological defects in
superstring theories have led to a renewed interest in cosmic
(super)strings. In particular, a variant of their formation mechanism has
been proposed in the framework of brane inflation \cite{Sara02}. String-like
defects also appear in a number of condensed matter systems, including
liquid crystals and graphene-made structures.

In the simplest theoretical model, the cosmic string is described by a
planar angle deficit with the background geometry being locally flat except
on the top of the string where it has a delta shaped curvature tensor. The
corresponding non-trivial topology induces nonzero vacuum expectation values
(VEVs) for physical observables. Specifically, the VEV of the
energy-momentum tensor associated with various fields has been developed by
many authors \cite{Hell86}-\cite{Beze06b}. Moreover, considering a magnetic
flux running along the strings, there appear additional contributions to the
corresponding vacuum polarization effects for charged fields \cite{Dowk87},%
\cite{charged1}-\cite{Site12}. The presence of a magnetic flux induces also
vacuum current densities. This phenomenon was analyzed for massless \cite%
{Sira} and massive \cite{Yu} scalar fields. It has been shown that an
azimuthal vacuum current appears if the ratio of the magnetic flux by the
quantum one has a nonzero fractional part. The analysis of the induced
fermionic currents in higher-dimensional cosmic string spacetime in the
presence of a magnetic flux have been developed in \cite{Mello10}. The
fermionic current induced by a magnetic flux in (2+1)-dimensional conical
spacetime and in the presence of a circular boundary has also been analyzed
\cite{Saha10}.

In general, the analysis of quantum effects for matter fields in a cosmic
string spacetime, consider this defect in a flat background geometry. For a
cosmic string in a curved background, quantum effects associated with a
scalar field have been discussed in \cite{Davi88} for special values of the
planar angle deficit. The vacuum polarization in Schwarzschild space-time
threaded by an infinite straight cosmic string is investigated in \cite%
{Otte10}. In recent publications we have investigated the vacuum
polarization effects for massive scalar \cite{Beze09} and fermionic \cite%
{Beze10} fields, induced by a cosmic string in dS spacetime. It has been
shown that for massive quantum fields the background gravitational field
essentially changes the behavior of the vacuum densities at distances from
the string larger than the dS curvature radius, when compared with the case
of the string in Minkowski spacetime. Depending on the specific value of the
mass, at large distances two regimes are realized with monotonic and
oscillatory behavior of the VEVs. Similar analysis for vacuum polarization
effects, induced by a cosmic string in anti-de Sitter spacetime, have been
developed in \cite{Beze12} and \cite{Beze13} for massive scalar and
fermionic fields, respectively.

The choice of dS spacetime as the background geometry in the present paper
is motivated by several reasons. First of all, this spacetime is a maximally
symmetric solution of the Einstein equation in the presence of a positive
cosmological constant and, as a consequence of high degree of symmetry, a
large number of physical problems are exactly solvable on its background. As
it will be shown below, this is the case for the problem under
consideration. The importance of dS spacetime as a gravitational background
has essentially increased after the appearance of the inflationary scenario
for the expansion of the universe at early stages. Most versions of this
scenario assume a period of quasiexponential expansion in which the geometry
of the universe is approximated by a portion of dS spacetime. This gives a
natural solution to a number of problems in standard cosmology. In addition,
the quantum fluctuations in the inflaton field during the inflationary epoch
generate inhomogeneities that are seeds for the formation of the large scale
cosmic structures. More recently, astronomical observations of high-redshift
supernovae, galaxy clusters, and the cosmic microwave background have
indicated that at present the universe is accelerating and can be well
approximated by the Friedmann-Robertson-Walker cosmological model with the
energy dominated by a positive cosmological constant-type source. If the
universe is going to accelerate forever, this model will lead asymptotically
to a dS spacetime as a future attractor for the dynamics of the universe.

The second type of the topological effects we shall consider here is induced
by the compactification of the spatial dimension along the cosmic string
axis. The compact spatial dimensions are an inherent feature of most
high-energy theories of fundamental physics, including supergravity and
superstring theories. An interesting application of the field theoretical
models with compact dimensions recently appeared in nanophysics. The
long-wavelength description of the electronic states in graphene can be
formulated in terms of the Dirac-like theory in three-dimensional spacetime
with the Fermi velocity playing the role of speed of light (see, e.g., \cite%
{Cast09}). In graphene-made structures, like cylindrical and toroidal carbon
nanotubes, the background geometry for the corresponding field theory
contains one or two compact dimensions. In quantum field theory, the
periodicity conditions imposed on the field operator along compact
dimensions modify the spectrum for the normal modes and as a result of this
the VEVs of physical observables are changed. Recently the analysis of the
induced fermionic current and the VEV of the energy-momentum tensor in a
compactified cosmic string spacetime in the presence of magnetic flux
running along the string, have been developed in \cite{Mello13,Bellu14}. The
VEV of the fermionic current in spacetimes with an arbitrary number of
toroidally compactified spatial dimensions and in the presence of a constant
gauge has been investigated in \cite{Bell10FC}. Furthermore, the combined
effects of topology and the gravitational field on the VEVs of the current
density for charged scalar and fermionic fields in the background of dS
spacetime with an arbitrary number of toroidally compactified spatial
dimensions is considered in \cite{Bell12}. The finite temperature effects on
the current densities for scalar and fermionic fields in topologically
nontrivial spaces have been discussed in \cite{Beze13SC,Bell14b}.

The present paper is organized as follows. In section \ref{sec2} we describe
the background geometry and construct the complete set of normalized
positive- and negative-energy fermionic mode functions obeying a
quasiperiodic boundary condition with an arbitrary phase along the string
axis. In addition, we assume the presence of a constant gauge field. In
section \ref{sec3}, by using the mode-summation method, we first show that
the VEVs for the charge density and the radial current vanish. Then we
evaluate the renormalized VEV of the azimuthal current density induced by a
magnetic flux running along the string axis. It is decomposed into two
parts: the first one corresponds to the geometry of a cosmic string in dS
spacetime without compactification and the second one is induced by the
compactification of the spatial dimension parallel to the string. The VEV of
the axial current density is investigated in section \ref{sec4}. This VEV is
a purely topological effect induced by the compactification and vanishes in
the geometry of a straight cosmic string. The most relevant conclusions of
the paper are summarized in section \ref{conc}. Throughout the paper we use
the units with $G=\hbar =c=1$.

\section{Geometry of the problem and the fermionic modes}

\label{sec2}

The main objective of this section is to present the geometry of the
spacetime, where we develop our analysis and also to obtain the complete set
of solutions of Dirac equation in this background. So we first write the
line element, in cylindrical coordinates, corresponding to a cosmic string
along the $z$-axis in dS spacetime
\begin{equation}
ds^{2}=g_{\mu \nu }dx^{\mu }dx^{\nu }=dt^{2}-e^{2t/\alpha }\left(
dr^{2}+r^{2}d\phi ^{2}+dz{}^{2}\right) \ ,  \label{ds21}
\end{equation}%
where $r\geqslant 0$, $t\in (-\infty ,+\infty )$ and $0\leqslant \phi
\leqslant \phi _{0}$, being $\phi _{0}=2\pi /q$. The parameter $q$, bigger
than unity, codifies the presence of the cosmic string. Additionally we
shall assume that the direction along the $z$-axis is compactified to a
circle with the length $L$: $0\leqslant z\leqslant L$. The parameter $\alpha
$ in (\ref{ds21}) is related to the cosmological constant $\Lambda $ and the
scalar curvature $R$ by the expressions $\Lambda =3\alpha ^{-2}$ and $%
R=12\alpha ^{-2}$.

In addition to the synchronous time coordinate $t$, we introduce the
conformal time $\tau $\ according to
\begin{equation}
\tau =-\alpha e^{-t/\alpha }\ ,\ -\infty <\ \tau \ <\ 0\ .  \label{tau}
\end{equation}%
In terms of this coordinate, the line element (\ref{ds21}) is confromally
related to the geometry of a cosmic string in Minkowski bulk, with the
conformal factor $(\alpha /\tau )^{2}$:%
\begin{equation}
ds^{2}=(\alpha /\tau )^{2}\left( d\tau ^{2}-dr^{2}-r^{2}d\phi
^{2}-dz{}^{2}\right) .  \label{ds21b}
\end{equation}

By the coordinate transformation%
\begin{equation}
t=-\alpha \ln f(t_{s},r_{s}),\;r=r_{s}f(t_{s},r_{s})\sin \theta
,\;z=r_{s}f(t_{s},r_{s})\cos \theta ,  \label{coordtrans}
\end{equation}%
and $\phi =\varphi /q$, with the function $f(t_{s},r_{s})=e^{-t_{s}/\alpha }/%
\sqrt{1-r_{s}^{2}/\alpha ^{2}}$, the line element (\ref{ds21}) is presented
in the static form
\begin{equation}
ds^{2}=(1-r_{s}^{2}/\alpha ^{2})dt_{s}^{2}-\frac{dr_{s}^{2}}{%
1-r_{s}^{2}/\alpha ^{2}}-r_{s}^{2}(d\theta ^{2}+q^{-2}\sin ^{2}\theta
d\varphi ^{2}).  \label{ds2st}
\end{equation}%
This line element has been previously discussed in \cite{Ghez02}. It is
shown that, to leading order in the gravitational coupling, the effect of
the vortex on de Sitter spacetime is described by (\ref{ds2st}).

The dynamics of a massive spinor field in curved spacetime in the presence
of a four-vector potential, $A_{\mu }$, is governed by the Dirac equation
\begin{equation}
i\gamma ^{\mu }\left( \nabla _{\mu }+ieA_{\mu }~\right) \psi -m\psi =0\ ,%
\mathrm{~}\nabla _{\mu }=\partial _{\mu }+\Gamma _{\mu }\ .  \label{Direq}
\end{equation}%
Here, $\gamma ^{\mu }$ represents the Dirac matrix in curved spacetime and $%
\Gamma _{\mu }$ the spin connection. Both are expressed in terms of the flat
space Dirac matrices, $\gamma ^{(a)}$, by the relations
\begin{equation}
\gamma ^{\mu }=e_{(a)}^{\mu }\gamma ^{(a)},\;\Gamma _{\mu }=\frac{1}{4}%
\gamma ^{(a)}\gamma ^{(b)}e_{(a)}^{\nu }e_{(b)\nu ;\mu }\ ,  \label{Gammamu}
\end{equation}%
where the semicolon stands for the standard covariant derivative for vector
fields. In (\ref{Gammamu}), $e_{(a)}^{\mu }$ is the tetrad basis satisfying
the relation $e_{(a)}^{\mu }e_{(b)}^{\nu }\eta ^{ab}=g^{\mu \nu }$, with $%
\eta ^{ab}$ being the Minkowski spacetime metric tensor.

We assume that along the compact $z$-dimension the fermionic field obeys the
quasiperiodicity condition as shown below:
\begin{equation}
\psi (t,r,\phi ,z+L)=e^{2\pi i\beta }\psi (t,r,\phi ,z)\ .  \label{Period}
\end{equation}%
In the above equation, $\beta $ is a constant phase defined in the interval $%
[0,\ 1]$. The special cases $\beta =0$ and $\beta =1/2$ correspond to the
periodic and antiperiodic boundary conditions (untwisted and twisted fields,
respectively). For the rotation around the $z$-axis we shall use the
periodic boundary condition
\begin{equation}
\psi (t,r,\phi +\phi _{0},z)=\psi (t,r,\phi ,z).  \label{Period2}
\end{equation}

For a constant vector potential, $A_{\mu }$, the latter may be excluded from
the field equation (\ref{Direq}) by the gauge transformation%
\begin{equation}
A_{\mu }^{\prime }=A_{\mu }+\partial _{\mu }\Lambda ,\;\psi ^{\prime
}=e^{-ie\Lambda }\psi ,  \label{Gauge}
\end{equation}%
with $\Lambda =-A_{\mu }x^{\mu }$. The new wave function obeys the equation
\begin{equation}
(i\gamma ^{\mu }\nabla _{\mu }-m)\psi ^{\prime }=0,  \label{Direq2}
\end{equation}%
and the periodicity conditions%
\begin{eqnarray}
\psi ^{\prime }(t,r,\phi +\phi _{0},z) &=&e^{2\pi ia}\psi ^{\prime
}(t,r,\phi ,z),  \label{PerCond1} \\
\psi ^{\prime }(t,r,\phi ,z+L) &=&e^{2\pi i\tilde{\beta}}\psi ^{\prime
}(t,r,\phi ,z),  \label{PerCond2}
\end{eqnarray}%
with the notations%
\begin{equation}
a=eA_{2}/q,\;\tilde{\beta}=\beta +eA_{3}L/2\pi .  \label{abet}
\end{equation}%
Note that the physical components $A_{\phi }$ and $A_{z}$ of the vector
potential are related to the covariant components $A_{2}$ and $A_{3}$ by $%
A_{\phi }=-A_{2}/r$ and $A_{z}=-A_{3}$. The parameters in the phases of the
periodicity conditions can be expressed in terms of the magnetic flux along
the string axis, $\Phi _{2}=-A_{2}\phi _{0}$, and flux enclosed by the $z$%
-axis, $\Phi _{3}=-A_{3}L$, by the formulas%
\begin{equation}
a=-\Phi _{2}/\Phi _{0},\;\tilde{\beta}=\beta -\Phi _{3}/\Phi _{0},
\label{abet2}
\end{equation}%
with $\Phi _{0}=2\pi /e$ being the flux quantum. In what follows we will
work in terms of the gauge transformed field $\psi ^{\prime }$ omitting the
prime. The current density is invariant under the gauge transformation (\ref%
{Gauge}).

Our main interest in this paper is the evaluation of the VEV of the
fermionic current density, $j^{\mu }=e\bar{\psi}\gamma ^{\mu }\psi $. This
VEV is expressed in terms of the two-point function $S_{rs}^{(1)}(x,x^{%
\prime })=\langle 0|[\psi _{r}(x),\bar{\psi}_{s}(x^{\prime })]|0\rangle $,
where $r$ and $s$ are spinor indices and $|0\rangle $ is the vacuum state.
For the VEV one has%
\begin{equation}
\langle j^{\mu }(x)\rangle \equiv \langle 0|j^{\mu }(x)|0\rangle =-\frac{e}{2%
}\lim_{x^{\prime }\rightarrow x}\mathrm{Tr}(\gamma ^{\mu
}S^{(1)}(x,x^{\prime })).  \label{VEVj}
\end{equation}%
In quantum field theory on curved backgrounds the choice of the vacuum is
not unique (see, for example, \cite{Birr82}). In dS spacetime there exists a
one-parameter family of maximally symmetric quantum states. In what follows
we will assume that the field is prepared in the dS-invariant Bunch-Davies
vacuum state \cite{Bunc78}. In the class of dS-invariant quantum states, the
Bunch-Davies vacuum is the only one for which the ultraviolet behavior of
the two-point functions is the same as in Minkowski spacetime.

Let $\{\psi _{\sigma }^{(+)}(x),\psi _{\sigma }^{(-)}(x)\}$ be a complete
set of normalized solutions to the Dirac equation specified by the set of
quantum numbers $\sigma $. Note that the background geometry under
consideration is time-dependent and the energy is not conserved. However, we
will refer to the solutions $\psi _{\sigma }^{(+)}(x)$ and $\psi _{\sigma
}^{(-)}(x)$ as the positive- and negative-energy modes in the sense that in
the limit $\alpha \rightarrow \infty $ they reproduce the positive- and
negative-energy fermionic modes in Minkowski spacetime. Expanding the field
operator in terms of the complete set of fermionic modes, the following
mode-sum formula is obtained for the current density:
\begin{equation}
\langle j^{\mu }\rangle =\frac{e}{2}\sum_{\sigma }\left[ \bar{\psi}_{\sigma
}^{(-)}(x)\gamma ^{\mu }\psi _{\sigma }^{(-)}(x)-\bar{\psi}_{\sigma
}^{(+)}(x)\gamma ^{\mu }\psi _{\sigma }^{(+)}(x)\right] .  \label{current}
\end{equation}%
Consequently, in this evaluation we need the fermionic modes for the
geometry at hand.

In order to find the mode functions, we will take the flat space Dirac
matrices according to \cite{B-D}
\begin{equation}
\gamma ^{(0)}=\left(
\begin{array}{cc}
1 & 0 \\
0 & -1%
\end{array}%
\right) ,\;\gamma ^{(a)}=\left(
\begin{array}{cc}
0 & \sigma _{a} \\
-\sigma _{a} & 0%
\end{array}%
\right) ,  \label{gam0l}
\end{equation}%
where $a=1,2,3$, and $\sigma _{1},\sigma _{2},\sigma _{3}$ are the $2\times
2 $ Pauli matrices. The basis of tetrads corresponding to the line element (%
\ref{ds21}) may have the form%
\begin{equation}
e_{(a)}^{\mu }=e^{-t/\alpha }\left(
\begin{array}{cccc}
e^{t/\alpha } & 0 & 0 & 0 \\
0 & \cos (q\phi ) & -\sin (q\phi )/r & 0 \\
0 & \sin (q\phi ) & \cos (q\phi )/r & 0 \\
0 & 0 & 0 & 1%
\end{array}%
\right) .  \label{Tetrad}
\end{equation}%
For the curved space gamma matrices, in the coordinate system corresponding
to (\ref{ds21}), this choice leads to the representation%
\begin{equation}
\gamma ^{0}=\gamma ^{(0)},\;\gamma ^{l}=e^{-t/\alpha }\left(
\begin{array}{cc}
0 & \rho ^{l} \\
-\rho ^{l} & 0%
\end{array}%
\right) ,  \label{gaml}
\end{equation}%
with the $2\times 2$ matrices%
\begin{equation}
\rho ^{1}=\left(
\begin{array}{cc}
0 & e^{-iq\phi } \\
e^{iq\phi } & 0%
\end{array}%
\right) ,\;\rho ^{2}=-\frac{i}{r}\left(
\begin{array}{cc}
0 & e^{-iq\phi } \\
-e^{iq\phi } & 0%
\end{array}%
\right) ,  \label{betal}
\end{equation}%
and $\rho ^{3}=\sigma _{3}$. For the spin connection components\ one gets $%
\Gamma _{0}=0$ and
\begin{equation}
\Gamma _{l}=-\frac{1}{2\alpha }\gamma ^{0}\gamma _{l}+\frac{1-q}{2}\gamma
^{(1)}\gamma ^{(2)}\delta _{l}^{2},  \label{Gaml}
\end{equation}%
for $l=1,2,3$. This leads to the following expression for the combination
appearing in the Dirac equation (\ref{Direq}):
\begin{equation}
\gamma ^{\mu }\Gamma _{\mu }=\frac{3\gamma ^{0}}{2\alpha }+\frac{1-q}{2r}%
\gamma ^{1}.  \label{gamGam}
\end{equation}

The positive- and negative-energy mode functions obeying the periodicity
conditions (\ref{PerCond2}) can be found in a way similar to that we have
used in \cite{Beze10} for the geometry of a straight cosmic string in dS
spacetime in the absence of the magnetic flux. For the Bunch-Davies vacuum
state these functions are given by
\begin{equation}
\psi _{\sigma }^{(\pm )}(x)=C_{\sigma }^{(\pm )}\eta ^{2}e^{iq(j+a)\phi
+ikz}\left(
\begin{array}{c}
H_{1/2-im\alpha }^{(\lambda _{\pm })}(\gamma \eta )J_{\beta
_{1}}(pr)e^{-iq\phi /2} \\
\frac{isp\epsilon _{j}}{\gamma +sk}H_{1/2-im\alpha }^{(\lambda _{\pm
})}(\gamma \eta )J_{\beta _{2}}(pr)e^{iq\phi /2} \\
-isH_{-1/2-im\alpha }^{(\lambda _{\pm })}(\gamma \eta )J_{\beta
_{1}}(pr)e^{-iq\phi /2} \\
\frac{p\epsilon _{j}}{\gamma +sk}H_{-1/2-im\alpha }^{(\lambda _{\pm
})}(\gamma \eta )J_{\beta _{2}}(pr)e^{iq\phi /2}%
\end{array}%
\right) ,  \label{psisigma+}
\end{equation}%
where $\lambda _{+}=1$, $\lambda _{-}=2$, $j=\pm 1/2,\pm 3/2,\ldots $, $\eta
=|\tau |$, $s=\pm 1$. Moreover, $J_{\nu }(x)$ and $H_{\nu }^{(1,2)}(x)$ are
the Bessel and Hankel functions, respectively, and
\begin{equation}
\gamma =\sqrt{{k}^{2}+p^{2}},\;0\leqslant p<\infty .  \label{gamma}
\end{equation}%
In (\ref{psisigma+}), we have defined
\begin{eqnarray}
\beta _{1} &=&q|j+a|-\epsilon _{j}/2,  \notag \\
\beta _{2} &=&q|j+a|+\epsilon _{j}/2,  \label{betajn}
\end{eqnarray}%
with $\epsilon _{j}=1$ for $j>-a$ and $\epsilon _{j}=-1$ for $j<-a$. The
quantum number $j$ determines the eigenvalues of the projection of the total
momentum along the cosmic string and the quantum number $s$ corresponds to
the eigenvalue of%
\begin{equation*}
\hat{S}=\gamma ^{-1}\Sigma ^{n}\hat{p}_{n},\;\Sigma ^{n}=\left(
\begin{array}{cc}
\rho ^{n} & 0 \\
0 & \rho ^{n}%
\end{array}%
\right)
\end{equation*}
where $\hat{p}_{n}=-i\partial _{n}+\delta _{n}^{3}(q-1)\Sigma ^{3}/2$ with $%
n=1,2,3$.

The mode functions above are specified by the complete set of quantum
numbers $\sigma =(p,k,j,s)$. In addition, the functions (\ref{psisigma+})
obey the periodicity condition (\ref{PerCond1}). From the condition (\ref%
{PerCond2}) we find the eigenvalues for the quantum number $k$:
\begin{equation}
k=k_{l}=2\pi (l+{\tilde{\beta}})/L,  \label{keig}
\end{equation}%
with $l=0,\pm 1,\pm 2,\ldots $.

The coefficients $C_{\sigma }^{(\pm )}$ are determined by the
orthonormalization condition%
\begin{equation}
\int d^{3}x\sqrt{g^{(3)}}\psi _{\sigma }^{(\pm )\dagger }\psi _{\sigma
^{\prime }}^{(\pm )}=\delta _{\sigma \sigma ^{\prime }},  \label{norm}
\end{equation}%
where $g^{(3)}$ is the determinant of the spatial metric tensor
corresponding to the line element (\ref{ds21}). The delta symbol in the rhs
of (\ref{norm}) is understood as the Kronecker delta for the discrete
indices $(j,k_{l},s)$ and the Dirac delta function for the continuous one $%
p\in \lbrack 0,\ \infty )$. By using the Wronskian for the Hankel functions,
we find
\begin{equation}
|C_{\sigma }^{(\pm )}|^{2}=\frac{qpe^{\pm m\alpha \pi }}{16L\alpha ^{3}}%
(\gamma +sk).  \label{CN+}
\end{equation}%
Note that, if we write the parameter $a$, defined in (\ref{abet}), in the
form
\begin{equation}
a=n_{0}+a_{0}\ ,\ |a_{0}|<1/2\ ,  \label{alfa0}
\end{equation}%
where $n_{0}$ is an integer number, then, by shifting $j+n_{0}\rightarrow j$%
, we can see that the VEVs of physical observables depend on $a_{0}$ only.

As it is well known, in Minkowski spacetime, the theory of von Neumann
deficiency indices leads to a one-parameter (usually denoted by $\theta $)
family of allowed boundary conditions in the background of an Aharonov-Bohm
gauge field \cite{Sous89}. Additionally to the regular modes, these boundary
conditions, in general, allow normalizable irregular modes. A special case
of boundary conditions has been discussed in \cite{Bene00}, where the
Atiyah-Patodi-Singer type nonlocal boundary condition is imposed at a finite
radius, which is then taken to zero. Similar approach, with the MIT bag
boundary condition, has been used in \cite{Saha10,Bell11} for a
two-dimensional conical space with a circular boundary. In the geometry
under consideration there are no normalizable irregular modes for
\begin{equation}
|a_{0}|\leqslant (1-1/q)/2.  \label{Irno}
\end{equation}%
In the case $|a_{0}|>(1-1/q)/2$, the irregular mode corresponds to $j=-n_{0}-%
\mathrm{sgn\,}(a_{0})/2$. For the mode functions (\ref{psisigma+}) with this
value of the momentum, the boundary condition on the string axis is a
special case of one-parameter family of conditions with the parameter $%
\theta =\pi /2$. Note that with this value and for a massless field both
parity and chiral symmetry are conserved \cite{Site99}. The evaluation of
the VEV of the fermionic current for other boundary conditions on the string
axis is similar to that described below. The contribution of the regular
modes to the VEV is the same for all boundary conditions and the results
will differ by the parts related to the irregular modes.

\section{Charge, radial and azimuthal currents}

\label{sec3}

Having the complete set of mode functions (\ref{psisigma+}), we can evaluate
the VEV for the current density by making use of the mode-sum formula (\ref%
{current}) where now the summation is specified by%
\begin{equation}
\sum_{\sigma }=\int_{0}^{\infty }dp\ \sum_{l=-\infty }^{+\infty }\sum_{s=\pm
1}\sum_{j}\ ,  \label{Sumsig}
\end{equation}%
with%
\begin{equation}
\sum_{j}=\sum_{j=\pm 1/2,\pm 3/2,\cdots }.  \label{Sumj}
\end{equation}%
Of course, the expression in the rhs of (\ref{current}) is divergent and a
regularization with the subsequent renormalization is necessary. Here we
shall use a cutoff function to regularize without writing it explicitly. The
special form of this function will not be important for the further
discussion. An alternative way would be the point-splitting regularization
procedure which corresponds to the evaluation of the expression under the
sign of the limit in (\ref{VEVj}) for $x^{\prime }\neq x$. However, in this
case the calculations are more complicated.

First let us consider the charge density:
\begin{equation}
\langle j^{0}\rangle =-\frac{e}{2}\sum_{\sigma }\sum_{\chi =-,+}\chi \psi
_{\sigma }^{(\chi )\dagger }(x)\psi _{\sigma }^{(\chi )}(x).  \label{density}
\end{equation}%
Substituting the mode functions (\ref{psisigma+}) and using the relation
\cite{Abra64}
\begin{equation}
H_{\nu }^{(2)}(x)=\frac{2i}{\pi }e^{i\pi \nu /2}K_{\nu }(ix),  \label{Rel}
\end{equation}%
with $K_{\nu }(x)$ being the MacDonald function, we obtain%
\begin{eqnarray}
\langle j^{0}\rangle &=&-\frac{eq\eta ^{4}}{4\pi ^{2}L\alpha ^{3}}%
\int_{0}^{\infty }dp\,p\sum_{l=-\infty }^{+\infty }\gamma \sum_{j}\left[
J_{\beta _{1}}^{2}(pr)+J_{\beta _{2}}^{2}(pr)\right]  \notag \\
&&\times \sum_{\chi =-,+}\chi \left[ |K_{1/2+\chi im\alpha }(i\gamma \eta
)|^{2}+|K_{-1/2+\chi im\alpha }(i\gamma \eta )|^{2}\right] .
\label{density3}
\end{eqnarray}%
By taking into account that $K_{-\nu }(x)=K_{\nu }(x)$, we conclude that the
charge density vanishes.

For the VEV of the radial current density one has
\begin{equation}
\langle j^{1}\rangle =-\frac{e}{2}\sum_{\sigma }\sum_{\chi =-,+}\chi \psi
_{\sigma }^{(\chi )\dagger }(x)\gamma ^{0}\gamma ^{1}\psi _{\sigma }^{(\chi
)}(x).  \label{radial}
\end{equation}%
Substituting the corresponding gamma matrices and the fermionic mode
functions from (\ref{psisigma+}), it can be shown that all terms cancel and
the resulting radial current is also zero.

Now we turn to the azimuthal current which is given by the expression (\ref%
{current}) with $\mu =2$. Substituting (\ref{psisigma+}), we can see that
the positive- and negative-energy modes give the same contribution. By using
(\ref{Rel}), and after the summation over $s$, the VEV of the azimuthal
current is presented in the form
\begin{equation}
\langle j^{2}\rangle =-\frac{2eq\eta ^{5}}{\pi ^{2}L\alpha ^{4}r}%
\int_{0}^{\infty }dp\,p^{2}\sum_{j}\epsilon _{j}J_{\beta _{1}}(pr)J_{\beta
_{2}}(pr)\sum_{l=-\infty }^{+\infty }\mathrm{Re\,}\left[ K_{1/2+im\alpha
}(i\gamma \eta )K_{1/2+im\alpha }(-i\gamma \eta )\right] .  \label{FC2n}
\end{equation}%
In order to separate explicitly the topological part, for the summation over
$l$, we apply the Abel-Plana formula in the form \cite{SahaRev,Beze08}
\begin{eqnarray}
&&\frac{2\pi }{L}\sum_{l=-\infty }^{\infty
}g(k_{l})f(|k_{l}|)=\int_{0}^{\infty }du\,\left[ g(u)+g(-u)\right] f(u)
\notag \\
&&\qquad +i\int_{0}^{\infty }du\left[ f(iu)-f(-iu)\right] \sum_{\chi =\pm 1}%
\frac{g(i\chi u)}{e^{Lu+2\pi i\chi \tilde{\beta}}-1}\ ,  \label{FC3}
\end{eqnarray}%
choosing $g(u)=1$ and
\begin{equation}
f(u)=\mathrm{Re\,}[K_{1/2+im\alpha }(i\eta \sqrt{u^{2}+p^{2}}%
)K_{1/2+im\alpha }(-i\eta \sqrt{u^{2}+p^{2}})].  \label{FC4}
\end{equation}

The first term in the rhs of (\ref{FC3}) is responsible for the azimuthal
current in the cosmic string background without compactification, that will
be denoted below by $\left\langle j^{2}\right\rangle _{s}$. The second one
corresponds to the contribution due to the compactification of the string
along the $z$-axis, denoted by $\left\langle j^{2}\right\rangle _{c}$.
Therefore, the application of the summation formula (\ref{FC3}) allows us to
decompose the azimuthal current as
\begin{equation}
\left\langle j^{2}\right\rangle =\left\langle j^{2}\right\rangle
_{s}+\left\langle j^{2}\right\rangle _{c}.  \label{FC5}
\end{equation}%
As we shall see, the compactified part goes to zero in the limit $%
L\rightarrow \infty $.

We start the evaluation with the part corresponding to the geometry of a
straight cosmic string, $\left\langle j^{2}\right\rangle _{s}$. Using the
first term in the Abel-Plana formula, for this part we get
\begin{eqnarray}
\left\langle j^{2}\right\rangle _{s} &=&-\frac{2eq\eta ^{5}}{r\pi ^{3}\alpha
^{4}}\int_{0}^{\infty }dp\,p^{2}\int_{0}^{\infty }dk\sum_{j}\epsilon
_{j}J_{\beta _{1}}(pr)J_{\beta _{2}}(pr)  \notag \\
&&\times \mathrm{Re\,}\left[ K_{1/2+im\alpha }(i\gamma \eta )K_{1/2+im\alpha
}(-i\gamma \eta )\right] ,  \label{FC6}
\end{eqnarray}%
where $\gamma $ is given by the expression (\ref{gamma}). Replacing the
product of the MacDonald functions by the integral representation \cite%
{Wats44}
\begin{equation}
K_{\nu }(ix)K_{\nu }(-ix)=\int_{0}^{\infty }duu^{-1}\int_{0}^{\infty
}dy\cosh {(2\nu y)}\exp \left[ -2ux^{2}\sinh ^{2}y-1/(2u)\right] ,
\label{FC7}
\end{equation}%
the integral over $k$ is evaluated directly. Performing the integral over $p$
with the help of the formula below \cite{Gradshteyn}
\begin{equation}
\int_{0}^{\infty }dp\,p^{2}e^{-bp^{2}}J_{\beta _{1}}(pr)J_{\beta
_{2}}(pr)=r\epsilon _{j}\frac{e^{-r^{2}/2b}}{4b^{2}}\left[ I_{\beta
_{1}}(r^{2}/2b)-I_{\beta _{2}}(r^{2}/2b)\right] ,  \label{FC8}
\end{equation}%
we obtain%
\begin{eqnarray}
\left\langle j^{2}\right\rangle _{s} &=&-\frac{\sqrt{2}eq\eta ^{5}}{r^{5}\pi
^{5/2}\alpha ^{4}}\int_{0}^{\infty }dy\,\cosh y\cos (2m\alpha y)  \notag \\
&&\times \int_{0}^{\infty }dz\,z^{3/2}e^{-z\left[ 1+2(\eta /r)^{2}\sinh ^{2}y%
\right] }\mathcal{J}(q,a_{0},z),  \label{FC9}
\end{eqnarray}%
where we have introduced the notation%
\begin{equation}
\mathcal{J}(q,a_{0},z)=\sum_{j}\left[ I_{\beta _{1}}(z)-I_{\beta _{2}}(z)%
\right] .  \label{Jcal}
\end{equation}%
The integration over $y$ in (\ref{FC9}) can be done explicitly and one finds%
\begin{equation}
\left\langle j^{2}\right\rangle _{s}=-\frac{eq\alpha ^{-4}}{\sqrt{2}\pi
^{5/2}}\int_{0}^{\infty }dx\,x^{3/2}e^{x(1-r^{2}/\eta ^{2})}\mathcal{J}%
(q,a_{0},xr^{2}/\eta ^{2})\,\mathrm{Re\,}\left[ K_{1/2+im\alpha }(x)\right] ,
\label{FC13}
\end{equation}%
where we have introduced a new integration variable $x=z\eta ^{2}/r^{2}$.
The current density given by this formula is a periodic odd function of the
flux along the string axis with the period equal to the flux quantum. The
azimuthal current density, given by (\ref{FC13}), depends on the radial and
time coordinates through the ratio $r/\eta $. This property is a consequence
of the maximal symmetry of dS spacetime and the Bunch-Davies vacuum. By
taking into account that the combination $r_{p}=\alpha r/\eta $ is the
proper distance from the string, we see that $r/\eta $ is the proper
distance measured in units of the dS curvature scale $\alpha $.

For the further transformation of the expression in the rhs of (\ref{FC13}),
we follow the procedure used in \cite{Saha10} for the summation over $j$ in (%
\ref{Jcal}). Introducing the notation
\begin{equation}
\mathcal{I}(q,a_{0},z)=\sum_{j}I_{\beta _{1}}(z),  \label{seriesI0}
\end{equation}%
we can see that $\sum_{j}I_{\beta _{2}}(z)=\mathcal{I}(q,-a_{0},z)$. For the
series (\ref{seriesI0}) one has the integral representation
\begin{eqnarray}
&&\mathcal{I}(q,a_{0},z)=\frac{e^{z}}{q}-\frac{1}{\pi }\int_{0}^{\infty }dx%
\frac{e^{-z\cosh x}f(q,a_{0},x)}{\cosh (qx)-\cos (q\pi )}  \notag \\
&&\qquad +\frac{2}{q}\sum_{k=1}^{p}(-1)^{k}\cos [2\pi
k(a_{0}-1/(2q))]e^{z\cos (2\pi k/q)},  \label{seriesI3}
\end{eqnarray}%
in which $p$ is an integer defined by $2p<q<2p+2$ and
\begin{equation}
f(q,a_{0},x)=\sum_{\chi =-,+}\chi \cos \left[ q\pi \left( 1/2-\chi
a_{0}\right) \right] \cosh \left[ \left( qa_{0}+\chi q/2-1/2\right) x\right]
\ .  \label{fqualf1}
\end{equation}%
In the specific case $q=2p$, the term
\begin{equation}
-(-1)^{q/2}\frac{e^{-z}}{q}\sin {(q\pi a_{0})}  \label{AddJ}
\end{equation}%
should be added to the rhs of (\ref{seriesI3}). By taking into account (\ref%
{seriesI3}), for the function (\ref{Jcal}) one gets%
\begin{eqnarray}
\mathcal{J}(q,a_{0},z) &=&\frac{4}{\pi }\int_{0}^{\infty }dx\frac{e^{-z\cosh
(2x)}g(q,a_{0},2x)\cosh x}{\cosh (2qx)-\cos (q\pi )}  \notag \\
&&+\frac{4}{q}\sum_{k=1}^{p}(-1)^{k}\sin \left( \pi k/q\right) \sin \left(
2\pi ka_{0}\right) e^{z\cos (2\pi k/q)},  \label{Jcal2}
\end{eqnarray}%
where%
\begin{equation}
g(q,a_{0},x)=\sum_{\chi =-,+}\chi \cos \left[ q\pi \left( 1/2+\chi
a_{0}\right) \right] \cosh \left[ \left( 1/2-\chi a_{0}\right) qx\right] .
\label{gqa}
\end{equation}%
In the case $q=2p$, the term $-2(-1)^{q/2}(e^{-z}/q)\sin {(q\pi a_{0})}$
must be added to the rhs of (\ref{Jcal2}). As is seen, in the absence of a
magnetic flux along the string, $a_{0}=0$, one has $\mathcal{J}(q,0,z)=0$
and the azimuthal current density $\left\langle j^{2}\right\rangle _{s}$
vanishes which is the same result as for the flat space in the presence of
the cosmic string \cite{Mello13}.

The azimuthal current density in dS spacetime induced by the magnetic flux
is obtained as a special case with $q=1$. In this case, for the function in
the integrand of (\ref{Jcal2}) one has%
\begin{equation}
g(1,a_{0},2x)=-2\sin (\pi a_{0})\cosh (2a_{0}x)\cosh x,  \label{g1}
\end{equation}%
and the sum in the rhs of (\ref{Jcal2}) is absent. For the function $%
\mathcal{J}(q,a_{0},z)$ one gets
\begin{equation}
\mathcal{J}(1,a_{0},z)=-\frac{2}{\pi }\sin (\pi a_{0})\int_{0}^{\infty
}du\,e^{-z\cosh u}\cosh (a_{0}u).  \label{Jcalq1}
\end{equation}%
As a result, in the absence of the planar angle deficit the expression for
the azimuthal current density is simplified to
\begin{eqnarray}
\left\langle j^{2}\right\rangle _{s} &=&2\sqrt{2}\frac{e\sin (\pi a_{0})}{%
\pi ^{7/2}\alpha ^{4}}\int_{0}^{\infty }du\,\cosh (2a_{0}u)  \notag \\
&&\times \int_{0}^{\infty }dx\,x^{3/2}e^{x[1-2(r/\eta )^{2}\cosh ^{2}u]}\,%
\mathrm{Re\,}\left[ K_{1/2+im\alpha }(x)\right] .  \label{j2q1}
\end{eqnarray}

An equivalent expression for the azimuthal current in general case of $q$ is
obtained substituting (\ref{Jcal2}) into (\ref{FC9}) and integrating over $z$%
:%
\begin{eqnarray}
\left\langle j^{2}\right\rangle _{s} &=&-\frac{3e}{4\pi ^{2}(\alpha r/\eta
)^{4}}\int_{0}^{\infty }du\,\cos (2m\alpha \,\mathrm{arcsinh\,}(ur/\eta ))
\notag \\
&&\times \left[ \frac{q}{\pi }\int_{0}^{\infty }dx\frac{g(q,a_{0},2x)\cosh x%
}{\cosh (2qx)-\cos (q\pi )}\left( \cosh ^{2}x+u^{2}\right) ^{-5/2}\right.
\notag \\
&&\left. +\sum_{k=1}^{p}\frac{(-1)^{k}\sin \left( \pi k/q\right) \sin \left(
2\pi ka_{0}\right) }{\left[ \sin ^{2}(\pi k/q)+u^{2}\right] ^{5/2}}\right] ,
\label{FC9b}
\end{eqnarray}%
where we have made the change of variables $u=(\eta /r)\sinh y$. In the
absence of the conical defect one has $q=1$ and from (\ref{FC9b}) we find a
simpler formula
\begin{equation}
\left\langle j^{2}\right\rangle _{s}=\frac{3e\sin (\pi a_{0})}{4\pi
^{3}\alpha ^{4}}\int_{0}^{\infty }du\,\cosh (2a_{0}u)\int_{0}^{\infty }dx\,%
\frac{\cos (2m\alpha \,\mathrm{arcsinh\,}x)}{[x^{2}+(r/\eta )^{2}\cosh
^{2}u]^{5/2}}.  \label{j2bq1}
\end{equation}%
For a massless field, integrating over $u$, from the general formula (\ref%
{FC9b}) we find%
\begin{eqnarray}
\left\langle j^{2}\right\rangle _{s} &=&-\frac{e}{2\pi ^{2}(\alpha r/\eta
)^{4}}\left[ \sum_{k=1}^{p}\frac{(-1)^{k}\sin \left( 2\pi ka_{0}\right) }{%
\sin ^{3}(\pi k/q)}\right.  \notag \\
&&+\left. \frac{q}{\pi }\int_{0}^{\infty }dx\,\frac{g(q,a_{0},2x)\cosh ^{-3}x%
}{\cosh (2qx)-\cos (q\pi )}\right] .  \label{FC16}
\end{eqnarray}%
The above result shows that $\left\langle j^{2}\right\rangle _{s}$ is
conformally related to the corresponding induced current in pure cosmic
string spacetime \cite{Mello13}, with the conformal factor $\left( \eta
/\alpha \right) ^{4}$, as expected.

In the region near the string, $r/\eta \ll 1$, the dominant contribution to (%
\ref{FC13}) comes from large $x$ and we can use the asymptotic expression of
the MacDonald function for large arguments. The leading term is independent
of the mass and it reduces to (\ref{FC16}) which diverges on the string with
the inverse fourth power of the proper distance. At large distances from the
string, $r/\eta \gg 1$, in (\ref{FC13}) we use the asymptotic form of the
MacDonald function for small values of the argument. The leading term in the
asymptotic expansion of the azimuthal current behaves as
\begin{equation}
\left\langle j^{2}\right\rangle _{s}\propto \frac{\cos \left[ 2m\alpha \ln
(r/\eta )+\theta \right] }{(\alpha r/\eta )^{4}},  \label{j2sLarge}
\end{equation}%
with a phase $\theta $ depending on the parameters $m\alpha $, $q$ and $a_{0}
$. As a result, at large distances the azimuthal current in the geometry of
a straight cosmic string damps oscillatory with the amplitude decaying as
the inverse fourth power of the proper distance from the string. The
oscillation frequency increases with increasing mass. In the region under
consideration the influence of the gravitational field on the current
density is essential. The behavior of the current density in dS spacetime,
described by (\ref{j2sLarge}), is crucially different from that for the
string in Minkowski bulk. In the latter case, at large distances from the
string the current density for a massive field is suppressed exponentially,
by the factor $e^{-2mr\sin (\pi /q)}$ for $q\geqslant 2$ and by the factor $%
e^{-2mr}$ for $q<2$.

\begin{figure}[tbph]
\begin{center}
\begin{tabular}{cc}
\epsfig{figure=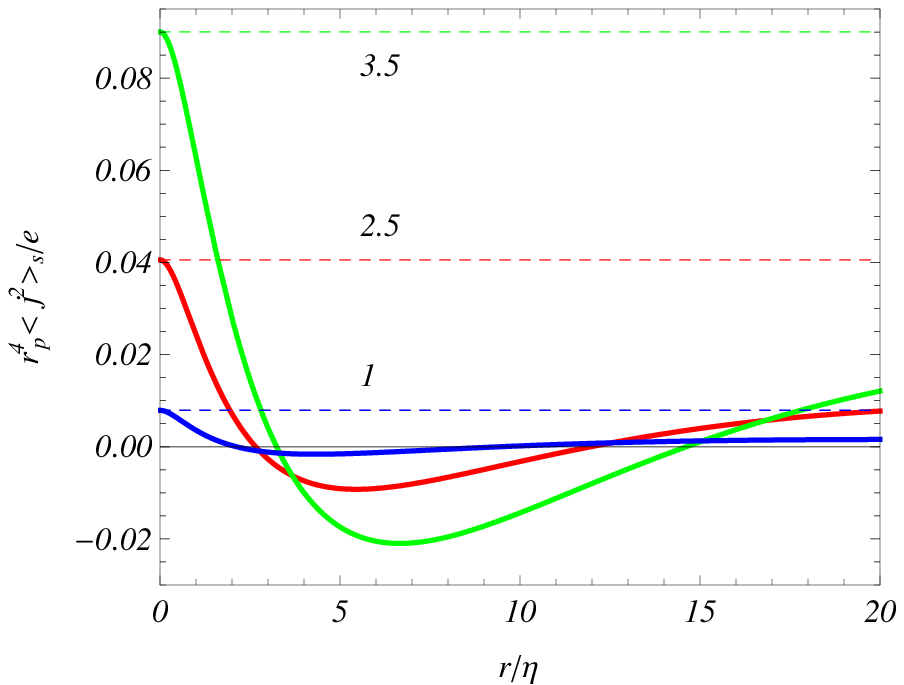,width=7.cm,height=5.5cm} & \quad %
\epsfig{figure=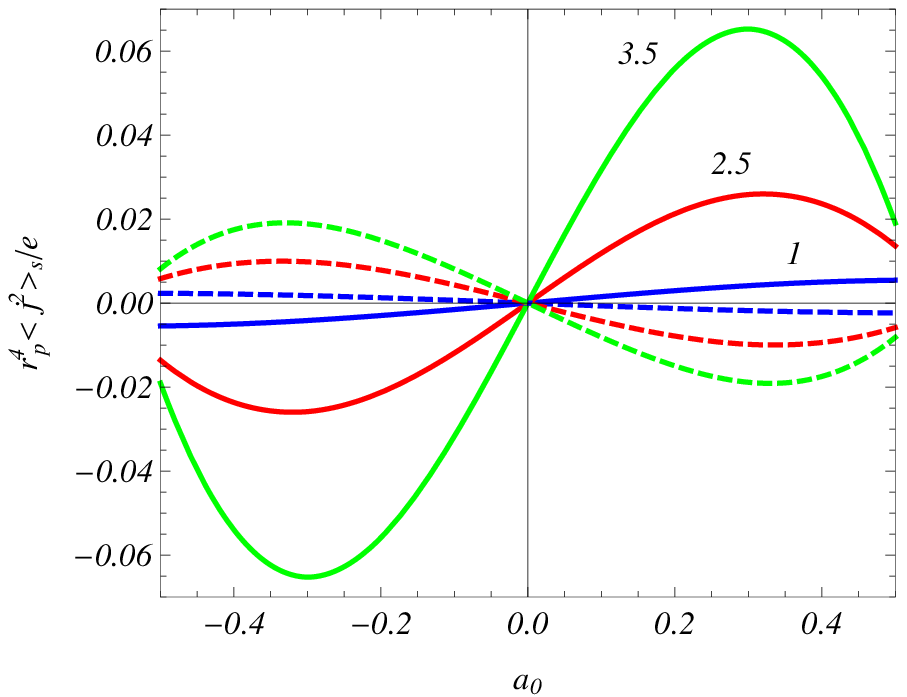,width=7.cm,height=5.5cm}%
\end{tabular}%
\end{center}
\caption{The azimuthal current in the geometry of a straight cosmic string,
multiplied by $r_{p}^{4}$, as a function of the distance from the string
(left panel) and as a function of the parameter $a_{0}$ (right panel) for
separate values of $q$ (numbers near the curves). On the left panel the
graphs are plotted for $a_{0}=1/4$. The full curves correspond to a massive
field with $m\protect\alpha =1$ and the dashed curves are for a massless
field. On the right panel the graphs are plotted for fixed values of $r/%
\protect\eta =1$ (full curves) and $r/\protect\eta =5$ (dashed curves).}
\label{fig1}
\end{figure}

In the left panel of figure \ref{fig1} we have plotted the quantity $%
r_{p}^{4}\left\langle j^{2}\right\rangle _{s}/e$, with $r_{p}=\alpha r/\eta $
being the proper distance from the string, as a function of the ratio $%
r/\eta $ (proper distance measured in units of $\alpha $) for $a_{0}=1/4$
and for separate values of the parameter $q$ (the numbers near the curves).
The full curves correspond to a massive field with $m\alpha =1$ and the
dashed lines are for a massless field. In the latter case the combination $%
r_{p}^{4}\left\langle j^{2}\right\rangle _{s}$ does not depend on $r/\eta $.
From the asymptotic analysis given above it follows that $%
r_{p}^{4}\left\langle j^{2}\right\rangle _{s}|_{r=0}=r_{p}^{4}\left\langle
j^{2}\right\rangle _{s}|_{m=0}$ which is also seen from the graphs. For
large values of $r/\eta $ and for a massive field we see the characteristic
oscillations described by (\ref{j2sLarge}). In the right panel of figure \ref%
{fig1} the quantity $r_{p}^{4}\left\langle j^{2}\right\rangle _{s}/e$ is
displayed as a function of the parameter $a_{0}$ for fixed values $r/\eta =1$
(full curves) and $r/\eta =5$ (dashed curves). Again, the numbers near the
full curves correspond to the values of $q$ and we have taken $m\alpha =1$.
For the dashed curves the same values of $q$ are used and $|\left\langle
j^{2}\right\rangle _{s}|$ increases with increasing $q$.

Now, we turn to the part in the azimuthal current density induced by the
compactification of the string along its axis. Using the second term in the
rhs of the Abel-Plana formula, from (\ref{FC2n}) we obtain%
\begin{eqnarray}
\langle j^{2}\rangle _{c} &=&\frac{eq\eta ^{5}}{\pi ^{3}\alpha ^{4}r}%
\int_{0}^{\infty }dp\,p^{2}\sum_{j}\epsilon _{j}J_{\beta _{1}}(pr)J_{\beta
_{2}}(pr)\int_{p}^{\infty }dk\,\sum_{\chi =\pm 1}\frac{1}{e^{Lk+2\pi i\chi
\tilde{\beta}}-1}  \notag \\
&&\times \mathrm{Im\,}\left\{ K_{1/2+im\alpha }(\eta \lambda )\left[
K_{1/2+im\alpha }(e^{\pi i}\eta \lambda )-K_{1/2+im\alpha }(e^{-\pi i}\eta
\lambda )\right] \right\} .  \label{FC18}
\end{eqnarray}%
where $\lambda =\sqrt{k^{2}-p^{2}}$. By employing the relation \cite{Abra64}
\begin{equation}
K_{\nu }(e^{im\pi }z)=e^{-im\nu \pi }K_{\nu }(z)-i\pi \frac{\sin (m\nu \pi )%
}{\sin (\nu \pi )}I_{\nu }(z),  \label{FC20}
\end{equation}%
we can see that
\begin{equation}
K_{\nu }(e^{i\pi }z)-K_{\nu }(e^{-i\pi }z)=-\pi i\left[ I_{-\nu }(z)+I_{\nu
}(z)\right] .  \label{Rel3}
\end{equation}%
Hence, for the topological part we obtain%
\begin{eqnarray}
\langle j^{2}\rangle _{c} &=&-\frac{2eq\eta ^{5}}{\pi ^{2}\alpha ^{4}r}%
\sum_{l=1}^{\infty }\cos (2\pi \tilde{\beta}l)\int_{0}^{\infty
}dp\,p^{2}\sum_{j}\epsilon _{j}J_{\beta _{1}}(pr)J_{\beta
_{2}}(pr)\int_{0}^{\infty }d\lambda \,\lambda  \notag \\
&&\times \frac{e^{-lL\sqrt{\lambda ^{2}+p^{2}}}}{\sqrt{\lambda ^{2}+p^{2}}}%
\mathrm{Re\,}\left\{ K_{1/2+im\alpha }(\eta \lambda )\left[ I_{-1/2-im\alpha
}(\eta \lambda )+I_{1/2+im\alpha }(\eta \lambda )\right] \right\} ,
\label{FC22}
\end{eqnarray}%
where we have used the expansion $\left( e^{u}-1\right)
^{-1}=\sum_{l=1}^{\infty }e^{-lu}$.

For the further transformation we employ the integral representation (see
also \cite{Bell14b})%
\begin{equation}
\frac{e^{-lL\sqrt{\lambda ^{2}+p^{2}}}}{\sqrt{\lambda ^{2}+p^{2}}}=\frac{2}{%
\sqrt{\pi }}\int_{0}^{\infty }ds\,e^{-(\lambda
^{2}+p^{2})s^{2}-l^{2}L^{2}/4s^{2}}.  \label{Rel4}
\end{equation}%
Substituting into (\ref{FC22}) and changing the order of integrations, the
integral over $p$ is evaluated by making use of (\ref{FC8}). For the
integral over $\lambda $ we use the formula%
\begin{equation}
\int_{0}^{\infty }d\lambda \,\lambda e^{-\lambda ^{2}s^{2}}K_{\nu }(\eta
\lambda )\left[ I_{-\nu }(\eta \lambda )+I_{\nu }(\eta \lambda )\right] =%
\frac{1}{2s^{2}}e^{\eta ^{2}/2s^{2}}K_{\nu }(\eta ^{2}/2s^{2}).  \label{Rel5}
\end{equation}%
The latter is obtained from the formula \cite{Gradshteyn}%
\begin{equation}
\int_{0}^{\infty }d\lambda \,\lambda e^{-\lambda ^{2}s^{2}}I_{\nu }^{2}(\eta
\lambda )=\frac{1}{2s^{2}}e^{\eta ^{2}/2s^{2}}I_{\nu }(\eta ^{2}/2s^{2}),
\label{Rel6}
\end{equation}%
by taking into account the relation between the functions $K_{\nu }(x)$ and $%
I_{\pm \nu }(x)$.

As a result, for the topological contribution in the azimuthal current
density we get%
\begin{eqnarray}
\langle j^{2}\rangle _{c} &=&-\frac{\sqrt{2}eq}{\pi ^{5/2}\alpha ^{4}}%
\sum_{l=1}^{\infty }\cos (2\pi \tilde{\beta}l)\int_{0}^{\infty
}dx\,x^{3/2}e^{x[1-r^{2}/\eta ^{2}-l^{2}L^{2}/(2\eta ^{2})]}  \notag \\
&&\times \mathcal{J}(q,a_{0},xr^{2}/\eta ^{2})\,\mathrm{Re\,}\left[
K_{1/2+im\alpha }(x)\right] ,  \label{FC27}
\end{eqnarray}%
with the new integration variable $x=\eta ^{2}/(2s^{2})$ and with the
function $\mathcal{J}(q,a_{0},z)$ given by (\ref{Jcal2}). As is seen from
this expression, the topological part in the azimuthal current density is a
periodic odd function of the magnetic flux along the string and a periodic
even function of the flux enclosed by the compactified dimension. In both
cases, the period is equal to the flux quantum. By taking into account the
expression (\ref{FC13}), the total azimuthal current is written in the form%
\begin{eqnarray}
\langle j^{2}\rangle &=&-\frac{\sqrt{2}eq}{\pi ^{5/2}\alpha ^{4}}%
\sideset{}{'}{\sum}_{l=0}^{\infty }\cos (2\pi \tilde{\beta}%
l)\int_{0}^{\infty }dx\,x^{3/2}e^{x[1-r^{2}/\eta ^{2}-l^{2}L^{2}/(2\eta
^{2})]}  \notag \\
&&\times \mathcal{J}(q,a_{0},xr^{2}/\eta ^{2})\,\mathrm{Re\,}\left[
K_{1/2+im\alpha }(x)\right] ,  \label{j2tot}
\end{eqnarray}%
where the prime on the sign of the summation means that the term $l=0$
should be taken with the coefficient 1/2. The latter corresponds to $\langle
j^{2}\rangle _{s}$. The azimuthal current density depends on $L$, $r$, and $%
\eta $ through the ratios $L/\eta $ and $r/\eta $ which are the proper
length of the compact dimension and proper distance from the string axis
measured in units of $\alpha $. Again, this feature is a consequence of the
maximal symmetry of dS spacetime. In the absence of the planar angle deficit
one has $q=1$ and, by using (\ref{Jcalq1}), the general formula (\ref{j2tot}%
) is reduced to the expression%
\begin{eqnarray}
\langle j^{2}\rangle &=&\frac{4\sqrt{2}e}{\pi ^{7/2}\alpha ^{4}}\sin (\pi
a_{0})\sideset{}{'}{\sum}_{l=0}^{\infty }\cos (2\pi \tilde{\beta}%
l)\int_{0}^{\infty }du\,\cosh (2a_{0}u)\,  \notag \\
&&\times \int_{0}^{\infty }dx\,x^{3/2}e^{x[1-2(r/\eta )^{2}\cosh
^{2}u-l^{2}L^{2}/(2\eta ^{2})]}\,\mathrm{Re\,}\left[ K_{1/2+im\alpha }(x)%
\right] .  \label{j2totq1}
\end{eqnarray}%
The latter presents the azimuthal current induced by an infinitely thin flux
tube along the compactified $z$-axis.

For a massless field, by taking into account that $K_{1/2}(x)=e^{-x}\sqrt{%
\pi /2x}$ and using (\ref{Jcal2}), the integration over $x$ in (\ref{j2tot})
is done explicitly and one finds%
\begin{eqnarray}
\langle j^{2}\rangle &=&-\frac{16\pi ^{-2}e}{(\alpha /\eta )^{4}}%
\sideset{}{'}{\sum}_{l=0}^{\infty }\cos (2\pi \tilde{\beta}l)\left[
\sum_{k=1}^{p}\frac{(-1)^{k}\sin \left( \pi k/q\right) \sin \left( 2\pi
ka_{0}\right) }{[4r^{2}\sin ^{2}(\pi k/q)+l^{2}L^{2}]^{2}}\right.  \notag \\
&&+\left. \frac{q}{\pi }\int_{0}^{\infty }dx\frac{g(q,a_{0},2x)\cosh x}{%
\cosh (2qx)-\cos (q\pi )}(4r^{2}\cosh ^{2}x+l^{2}L^{2})^{-2}\right] \,.
\label{Azimuthal13}
\end{eqnarray}%
In this case the azimuthal current is equal to $(\eta /\alpha )^{4}$ times
the one for the flat space in the presence of the compactified cosmic string
\cite{Mello13}.

Let us discuss the behavior of the topological part in the azimuthal current
in the asymptotic regions of the parameters. First we consider the region
near the cosmic string. From (\ref{Jcal}) it follows that in the limit $%
z\rightarrow 0$, to the leading order one has%
\begin{equation}
\mathcal{J}(q,a_{0},z)\approx -\frac{\mathrm{sgn\,}%
(a_{0})(z/2)^{q(1/2-|a_{0}|)-1/2}}{\Gamma (q(1/2-|a_{0}|)+1/2)}.
\label{JcalNear}
\end{equation}%
Substituting this into (\ref{FC27}), to the same order for the topological
contribution we get%
\begin{eqnarray}
\langle j^{2}\rangle _{c} &=&\frac{\mathrm{sgn\,}(a_{0})\sqrt{2}eq(r/\sqrt{2}%
\eta )^{q(1-2|a_{0}|)-1}}{\pi ^{5/2}\alpha ^{4}\Gamma (q(1/2-|a_{0}|)+1/2)}%
\sum_{l=1}^{\infty }\cos (2\pi \tilde{\beta}l)  \notag \\
&&\times \int_{0}^{\infty }dx\,x^{q(1/2-|a_{0}|)+1}e^{x[1-l^{2}L^{2}/(2\eta
^{2})]}\,\mathrm{Re\,}\left[ K_{1/2+im\alpha }(x)\right] .  \label{j2cnear}
\end{eqnarray}%
From here it follows that the topological part in the azimuthal current
vanishes on the string for $|a_{0}|<(1-1/q)/2$, is finite for $%
|a_{0}|=(1-1/q)/2$ and diverges for $|a_{0}|>(1-1/q)/2$. Note that in the
latter case the divergent contribution comes from the irregular mode. For $%
a_{0}\neq 0$, in the region near the string the total current is dominated
by the part $\left\langle j^{2}\right\rangle _{s}$.

For large values $r$, $r/\eta \gg 1$, the dominant contribution to the
integral in (\ref{j2tot}) comes from the values of $x$ in the region $%
x\lesssim \eta ^{2}/r^{2}$ and we replace the MacDonald function by its
asymptotic form for small values of the argument. At large distances, the
behavior of the azimuthal current density depends crucially on whether the
parameter $\tilde{\beta}$ is zero or not. For $0<\tilde{\beta}<1$, we use
the relation
\begin{equation}
\sideset{}{'}{\sum}_{l=0}^{\infty }\cos (2\pi \tilde{\beta}%
l)e^{-yl^{2}L^{2}/(2r^{2})}\approx \frac{r}{L}\sqrt{\frac{\pi }{2y}}%
e^{-2(\pi \sigma _{\beta }r/L)^{2}/y},  \label{Rel8}
\end{equation}%
valid for $L/r\ll 1$, with $\sigma _{\beta }=\min (\tilde{\beta},1-\tilde{%
\beta})$ and $y=xr^{2}/\eta ^{2}$. For $q\geqslant 2$ the main contribution
comes from the term $k=1$ in the expression (\ref{Jcal2}) for the function $%
\mathcal{J}(q,a_{0},z)$. The remaining integral is expressed in terms of the
MacDonald function with the large argument. By using the corresponding
asymptotic expression, to the leading order one finds%
\begin{eqnarray}
\langle j^{2}\rangle &\approx &\frac{2e\sigma _{\beta }\eta ^{2}/L^{2}}{%
\alpha ^{4}\left( r/\eta \right) ^{2}}\frac{\sin \left( 2\pi a_{0}\right) }{%
\sin (\pi /q)}\frac{e^{-4\pi \sigma _{\beta }r\sin (\pi /q)/L}}{\sqrt{\cosh
(\pi m\alpha )}}\,  \notag \\
&&\times \cos \left[ m\alpha \ln \left( \frac{2rL\sin (\pi /q)}{\pi \sigma
_{\beta }\eta ^{2}}\right) +\beta _{0}(m\alpha )\right] ,  \label{j2far}
\end{eqnarray}%
where $\beta _{0}$ is the phase of the function $\Gamma (1/2+im\alpha )$.
Hence, for $\tilde{\beta}\neq 0$ at large distances the topological part in
the current density is exponentially small. In the case $q<2$ the
suppression is stronger, by the factor $e^{-4\pi \sigma _{\beta }r/L}$.

At large distances, $r/\eta \gg 1$, and for $\tilde{\beta}=0$, in (\ref%
{j2tot}) the dominant contribution to the series comes from large $l$ and we
use the asymptotic formula
\begin{equation}
\sideset{}{'}{\sum}_{l=0}^{\infty }e^{-yl^{2}L^{2}/(2r^{2})}\approx \frac{r}{%
L}\sqrt{\frac{\pi }{2y}}.  \label{Rel9}
\end{equation}%
The MacDonald function is replaced by its asymptotic form for small values
of the argument. After the integration over $x$ we get%
\begin{equation}
\langle j^{2}\rangle \approx -\frac{eq\eta B_{q}(a_{0},m\alpha )}{\pi
^{3}\alpha ^{4}L\left( r/\eta \right) ^{3}}\cos [2m\alpha \ln (2r/\eta
)+\beta _{q}(a_{0},m\alpha )],  \label{j2far2}
\end{equation}%
where the functions are defined by the relation%
\begin{eqnarray}
&&B_{q}(a_{0},m\alpha )e^{i\beta _{q}(a_{0},m\alpha )}=\Gamma (1/2+im\alpha
)\Gamma (3/2-im\alpha )  \notag \\
&&\qquad \times \left\{ \int_{0}^{\infty }dx\frac{g(q,a_{0},2x)\left( \cosh
x\right) ^{2im\alpha -2}}{\cosh (2qx)-\cos (q\pi )}+\frac{\pi }{q}%
\sum_{k=1}^{p}\frac{(-1)^{k}\sin \left( 2\pi ka_{0}\right) }{[\sin (\pi
k/q)]^{2-2im\alpha }}\right\} .  \label{Bq}
\end{eqnarray}%
In this case the decay is of power-law for both massive and massless fields.
This is in clear contrast with the case of the cosmic string in flat
spacetime where the decay of the topological part of the current density for
massive fields is exponential.

For large values $L/\eta $ and for fixed $r/\eta $, i.e. $L\gg r$, the
dominant contribution in the integral of (\ref{FC27}) comes from the region
near the lower limit of the integration. Expanding in the integrand over $x$
and by taking into account (\ref{JcalNear}), to the leading order we find%
\begin{eqnarray}
\langle j^{2}\rangle _{c} &\approx &\frac{4eqB_{2}(m\alpha )}{\pi
^{5/2}(\alpha L/\eta )^{4}}\frac{\mathrm{sgn\,}(a_{0})(r/L)^{q(1-2|a_{0}|)-1}%
}{\Gamma (q(1/2-|a_{0}|)+1/2)}  \notag \\
&&\times \sum_{l=1}^{\infty }\frac{\cos (2\pi \tilde{\beta}l)}{%
l^{q(1-2|a_{0}|)+3}}\,\cos [2m\alpha \ln \left( lL/\eta \right) +\beta
_{2}(m\alpha )].  \label{j2largeL}
\end{eqnarray}%
where the functions $B_{2}(m\alpha )$ and $\beta _{2}(m\alpha )$ are defined
by the relation%
\begin{equation}
B_{2}(m\alpha )e^{i\beta _{2}(m\alpha )}=\Gamma (1/2+im\alpha )\Gamma
(q(1/2-|a_{0}|)+3/2-im\alpha ),  \label{B2}
\end{equation}%
with $B_{2}(m\alpha )$ being the modulus of the expression on the right.
Note in the geometry of a cosmic string on Minkowski bulk the VEV\ of the
azimuthal current density for large values of $L$ is suppressed
exponentially, by the factor $e^{-mL}$.

For small values of $L$ and for $0<\tilde{\beta}<1$, by using the relation (%
\ref{Rel8}), it can be seen that the current density $\langle j^{2}\rangle $
is suppressed by the factor $e^{-4(\pi \sigma _{\beta }r/L)\sin \left( \pi
/q\right) }$ for $q\geqslant 2$ and by the factor $e^{-4\pi \sigma _{\beta
}r/L}$ for $q<2$. For $\tilde{\beta}=0$, by taking into account (\ref{Rel9}%
), to the leading order we get%
\begin{equation}
\langle j^{2}\rangle \approx -\frac{eq\eta }{\pi ^{2}\alpha ^{4}L}%
\int_{0}^{\infty }dx\,xe^{x(1-r^{2}/\eta ^{2})}\mathcal{J}%
(q,a_{0},xr^{2}/\eta ^{2})\,\mathrm{Re\,}\left[ K_{1/2+im\alpha }(x)\right] .
\label{smallL}
\end{equation}%
Note that in this limit the topological part dominates in the azimuthal
current density.

\begin{figure}[tbph]
\begin{center}
\begin{tabular}{cc}
\epsfig{figure=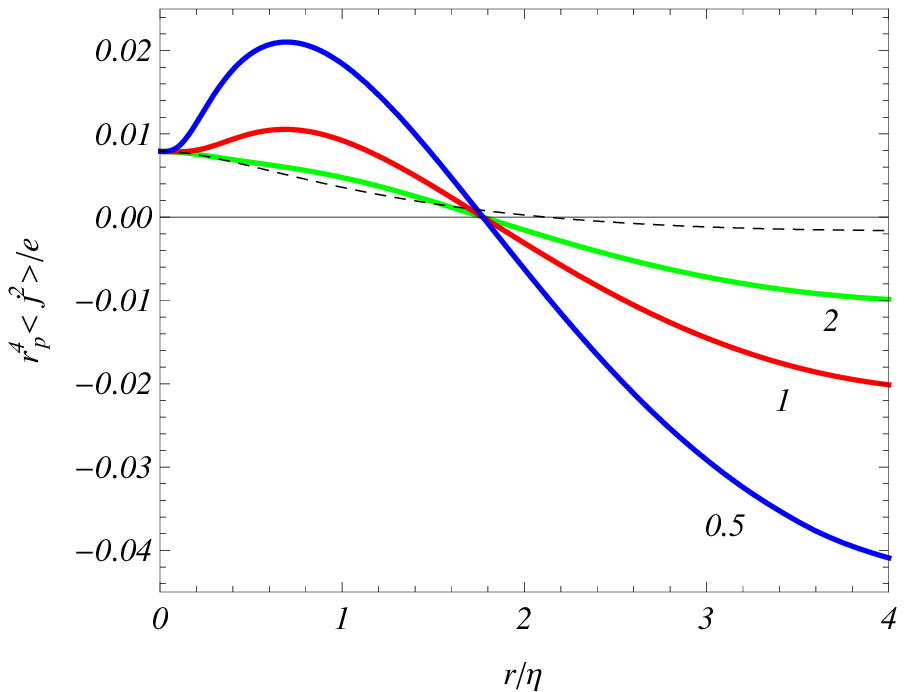,width=7.cm,height=5.5cm} & \quad %
\epsfig{figure=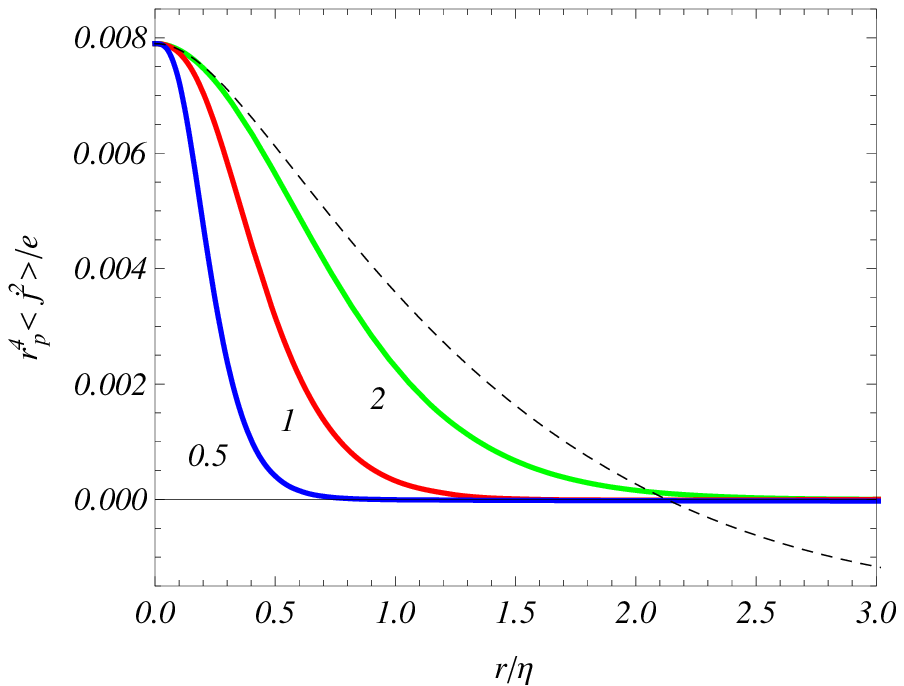,width=7.cm,height=5.5cm}%
\end{tabular}%
\end{center}
\caption{The azimuthal current density, multiplied by $r_{p}^{4}$, as a
function of the distance from the $z$-axis for $\tilde{\protect\beta}=0$
(left panel) and $\tilde{\protect\beta}=1/2$ (right panel). The values of
the other parameters are chosen as follows: $q=1$, $m\protect\alpha =1$, $%
a_{0}=1/4$. The dashed curves correspond to the azimuthal current in the
geometry of an uncompactified magnetic flux. The numbers near the curves
correspond to the values of $L/\protect\eta $.}
\label{fig2}
\end{figure}

In figure \ref{fig2}, we have presented the dependence of the azimuthal
current density, multiplied by $r_{p}^{4}$, on the proper distance from the $%
z$-axis measured in units of $\alpha $, $r/\eta $, for separate values of
the ratio $L/\eta $ (numbers near the curves). The dashed curves present the
corresponding quantity in the geometry of a uncompactified magnetic flux
along the $z$-axis, $r_{p}^{4}\left\langle j^{2}\right\rangle _{s}/e$. The
specific values of the parameters are chosen as follows: $q=1$, $m\alpha =1$%
, $a_{0}=1/4$. For the left panel $\tilde{\beta}=0$ and for the right one $%
\tilde{\beta}=1/2$. The features of the asymptotic analysis described above
are clearly seen in the numerical examples presented. In particular, for $%
\tilde{\beta}=0$ the topological part dominates for large $r/\eta $, whereas
for $\tilde{\beta}=1/2$ the current density is suppressed.

\begin{figure}[tbph]
\begin{center}
\begin{tabular}{cc}
\epsfig{figure=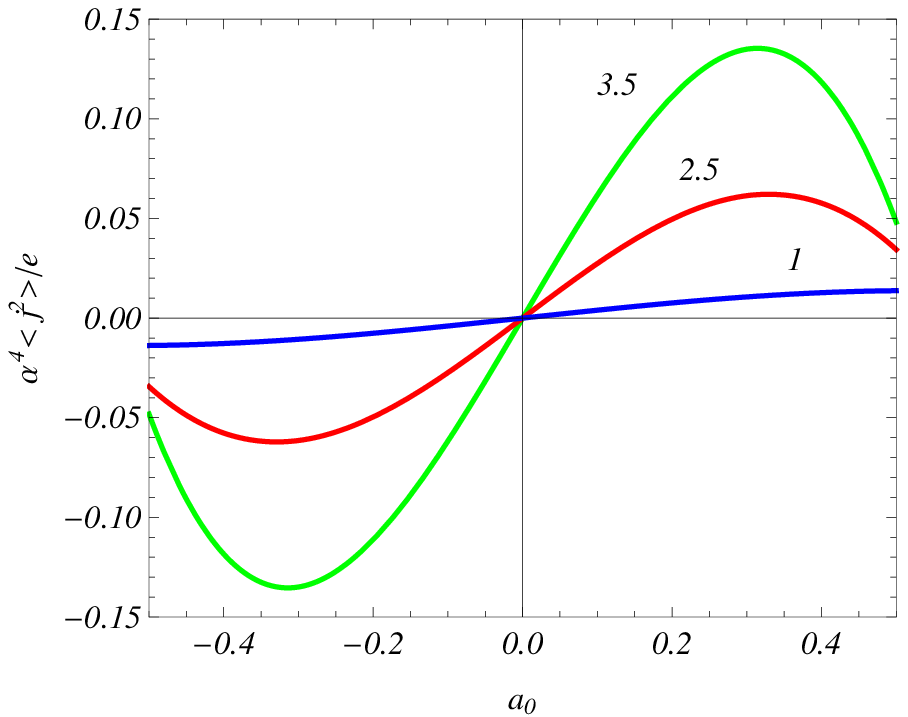,width=7.cm,height=5.5cm} & \quad %
\epsfig{figure=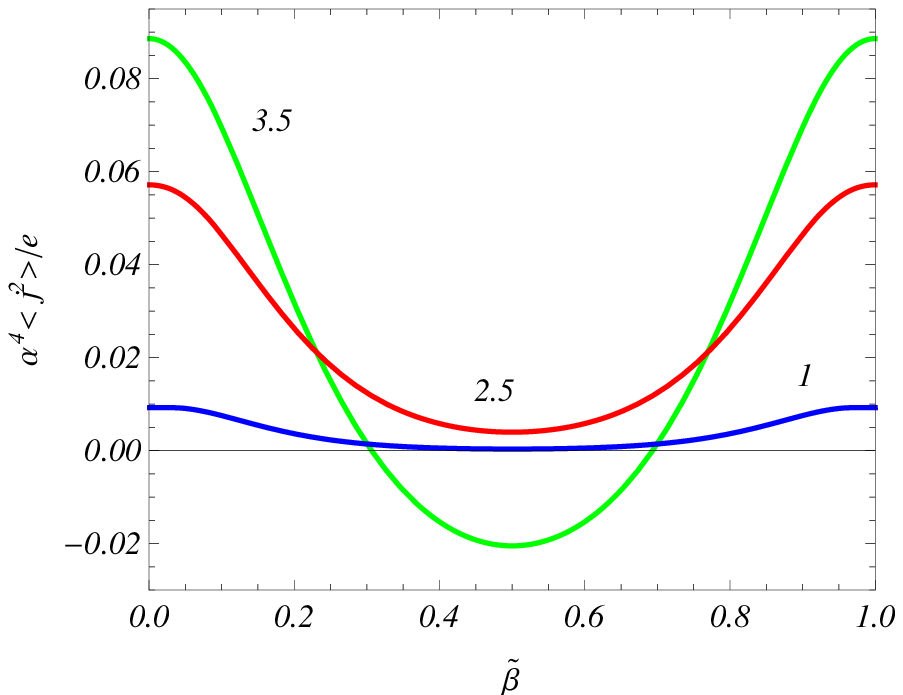,width=7.cm,height=5.5cm}%
\end{tabular}%
\end{center}
\caption{The azimuthal current density as a function of $a_{0}$ (left panel)
and $\tilde{\protect\beta}$ (right panel) for different values of the
parameter $q$ (numbers near the curves). The graphs are plotted for $m%
\protect\alpha =1$ and $L/\protect\eta =r/\protect\eta =1$. For the left
panel $\tilde{\protect\beta}=0$ and for the right one $a_{0}=1/4$.}
\label{fig3}
\end{figure}

In figure \ref{fig3}, the azimuthal current density is displayed as a
function of $a_{0}$ (left panel) and $\tilde{\beta}$ (right panel) for
different values of the parameter $q$ (numbers near the curves). The graphs
are plotted for the values of the parameters $m\alpha =1$ and $L/\eta
=r/\eta =1$. For the left panel we have taken $\tilde{\beta}=0$ and for the
right one $a_{0}=1/4$.

\section{Axial current}

\label{sec4}

In this section we investigate the axial component of the current density.
In order to do that we substitute the mode functions (\ref{psisigma+}) and
the relevant gamma matrices into (\ref{current}) with $\mu =3$. After some
intermediate steps the current density is presented in the form%
\begin{eqnarray}
\langle j^{3}\rangle  &=&-\frac{eq\eta ^{5}}{\pi ^{2}L\alpha ^{4}}%
\int_{0}^{\infty }dp\,p\,\sum_{j}[J_{\beta _{1}}^{2}(pr)+J_{\beta
_{2}}^{2}(pr)]  \notag \\
&&\times \sum_{l=-\infty }^{+\infty }k_{l}\,\mathrm{Re\,}\left[
K_{1/2+im\alpha }(i\gamma \eta )K_{1/2+im\alpha }(-i\gamma \eta )\right] ,
\label{Axial3}
\end{eqnarray}%
where we have replaced the Hankel functions by the MacDonald ones by using (%
\ref{Rel}). To perform the summation over $l$, we use again the Abel-Plana
formula (\ref{FC3}) as we did for $\langle j^{2}\rangle $, taking $g(u)=u$
and the expression given in Eq. (\ref{FC4}) for $f(u)$. The function $g(u)$
is an odd function which means that the first term on the rhs of the
Abel-Plana formula vanishes. Therefore, the axial current in the geometry of
a straight cosmic string vanishes. This result is in agreement with the one
obtained for the flat space in the presence of the cosmic string \cite%
{Mello13}.

The contribution to the axial current induced by the compactification is
given by the second term in the rhs of the summation formula (\ref{FC3}). It
is written in the form
\begin{eqnarray}
\langle j^{3}\rangle &=&-\frac{eq\eta ^{5}i}{2\pi ^{2}\alpha ^{4}}%
\int_{0}^{\infty }dp\,p\,\sum_{j}[J_{\beta _{1}}^{2}(pr)+J_{\beta
_{2}}^{2}(pr)]\int_{p}^{\infty }dk\,k\sum_{\chi =\pm 1}\frac{\chi }{%
e^{Lk+2\pi i\chi \tilde{\beta}}-1}  \notag \\
&&\times \mathrm{Re\,}\{K_{1/2+im\alpha }(\lambda \eta )\left[
I_{-1/2-im\alpha }(\lambda \eta )+I_{1/2+im\alpha }(\lambda \eta )\right] \},
\label{Axial3b}
\end{eqnarray}%
where $\lambda =\sqrt{k^{2}-p^{2}}$ and we have used the relation (\ref{Rel3}%
). For the further evaluation of this expression we first employ the
expansion $\left( e^{u}-1\right) ^{-1}=\sum_{l=1}^{\infty }e^{-lu}$ and then
change to the new integration variable $\lambda $. This leads to the
expression%
\begin{eqnarray}
\langle j^{3}\rangle &=&-\frac{eq\eta ^{5}}{\pi ^{2}\alpha ^{4}}%
\sum_{l=1}^{\infty }\sin (2\pi \tilde{\beta}l)\int_{0}^{\infty
}dp\,p\,\sum_{j}[J_{\beta _{1}}^{2}(pr)+J_{\beta
_{2}}^{2}(pr)]\int_{0}^{\infty }d\lambda \,\lambda  \notag \\
&&\times e^{-lL\sqrt{\lambda ^{2}+p^{2}}}\mathrm{Re\,}\{K_{1/2+im\alpha
}(\lambda \eta )\left[ I_{-1/2-im\alpha }(\lambda \eta )+I_{1/2+im\alpha
}(\lambda \eta )\right] \}.  \label{Axial3c}
\end{eqnarray}%
As the next step we use the integral representation%
\begin{equation}
e^{-lL\sqrt{\lambda ^{2}+p^{2}}}=\frac{lL}{\sqrt{\pi }}\int_{0}^{\infty
}ds\,s^{-2}e^{-(\lambda ^{2}+p^{2})s^{2}-l^{2}L^{2}/4s^{2}},  \label{IntRep2}
\end{equation}%
which is obtained from (\ref{Rel4}) by taking the derivative with respect to
$L$. Substituting this identity into (\ref{Axial3c}) and changing the order
of integrations, the integrals over $\lambda $ and $p$ are evaluated with
the help of formulas (\ref{Rel5}) and (\ref{Rel6}). Introducing the new
integration variable $x=\eta ^{2}/2s^{2}$, the final result is written as%
\begin{eqnarray}
\langle j^{3}\rangle &=&-\frac{\sqrt{2}eL}{\pi ^{5/2}\alpha ^{4}}%
\sum_{l=1}^{\infty }l\sin (2\pi \tilde{\beta}l)\int_{0}^{\infty
}dx\,x^{3/2}\,e^{x\left( 1-l^{2}L^{2}/2\eta ^{2}\right) }  \notag \\
&&\times \mathrm{Re\,}\left[ K_{1/2+im\alpha }(x)\right] \mathcal{F}%
(q,a_{0},xr^{2}/\eta ^{2}),  \label{Axial3d}
\end{eqnarray}%
with the function%
\begin{equation}
\mathcal{F}(q,a_{0},z)=\frac{q}{2}e^{-z}\left[ \mathcal{I}(q,a_{0},z)+%
\mathcal{I}(q,-a_{0},z)\right] .  \label{FCal}
\end{equation}%
For the latter, by using the representation (\ref{seriesI3}) for the
function $\mathcal{I}(q,a_{0},z)$, one gets%
\begin{eqnarray}
\mathcal{F}(q,a_{0},z) &=&1+2\sum_{k=1}^{p}(-1)^{k}\cos (2\pi ka_{0})\frac{%
\cos (\pi k/q)}{e^{2z\sin ^{2}(\pi k/q)}}  \notag \\
&&+\frac{2q}{\pi }\int_{0}^{\infty }dx\frac{e^{-2z\cosh
^{2}x}h(q,a_{0},2x)\sinh x}{\cosh (2qx)-\cos (q\pi )},  \label{FCal2}
\end{eqnarray}%
with the notation%
\begin{equation}
h(q,a_{0},x)=\sum_{\chi =\pm 1}\cos \left[ \left( 1/2+\chi a_{0}\right) q\pi %
\right] \sinh \left[ \left( 1/2-\chi a_{0}\right) qx\right] .  \label{h}
\end{equation}%
The axial current is an odd function of the parameter $\tilde{\beta}$ and an
even function of $a_{0}$.

The part in the current density coming from the first term in the right-hand
side of (\ref{FCal2}),%
\begin{equation}
\langle j^{3}\rangle _{0}=-\frac{\sqrt{2}eL}{\pi ^{5/2}\alpha ^{4}}%
\sum_{l=1}^{\infty }l\sin (2\pi \tilde{\beta}l)\int_{0}^{\infty
}dx\,x^{3/2}\,e^{x(1-l^{2}L^{2}/2\eta ^{2})}\mathrm{Re\,}\left[
K_{1/2+im\alpha }(x)\right] ,  \label{j30}
\end{equation}%
does not depend on the planar angle deficit and magnetic flux along the
string axis. It is a purely topological contribution and coincides with the
current density in dS spacetime with spatial topology $R^{2}\times S^{1}$ in
the absence of the string and of the magnetic flux along the $z$-axis. The
fermionic current in $(D+1)$-dimensional dS spacetime with topology $%
R^{p}\times (S^{1})^{D-p}$ has been investigated in \cite{Bell12}. By using
the integral representation
\begin{equation}
K_{\nu }(x)=2^{\nu -1}x^{-\nu }\int_{0}^{\infty }dy\,y^{\nu
-1}e^{-y-x^{2}/4y},  \label{KInt2}
\end{equation}%
and the formula (\ref{Rel5}), it can be seen that (\ref{j30}) coincides with
the result from \cite{Bell12} in the special case $D=3$, $p=2$. The part in
the axial current with the second and third terms on the right of (\ref%
{FCal2}) are induced by the presence of the string and of the magnetic flux
along its axis.

In the absence of the cosmic string one has $q=1$ and from (\ref{Axial3d})
we get
\begin{eqnarray}
\langle j^{3}\rangle &=&\langle j^{3}\rangle _{0}-\frac{2\sqrt{2}eL\sin
\left( a_{0}\pi \right) }{\pi ^{7/2}\alpha ^{4}}\sum_{l=1}^{\infty }l\sin
(2\pi \tilde{\beta}l)\int_{0}^{\infty }du\,\sinh (2a_{0}u)  \notag \\
&&\times \tanh (u)\int_{0}^{\infty }dx\,x^{3/2}\,e^{x[1-2(r/\eta )^{2}\cosh
^{2}u-l^{2}L^{2}/2\eta ^{2}]}\mathrm{Re\,}\left[ K_{1/2+im\alpha }(x)\right]
.  \label{j3q1}
\end{eqnarray}%
The second term on the rhs of this formula is induced by infinitely thin
magnetic flux running along the compactified $z$-axis.

For a massless Dirac field the MacDonald function in the general formula (%
\ref{Axial3d}) is expressed in terms of the exponential function.
Substituting (\ref{FCal2}) into (\ref{Axial3d}), after the integration over $%
z$ we find%
\begin{eqnarray}
\langle j^{3}\rangle &=&-\frac{4e(\eta /\alpha )^{4}}{\pi ^{2}L^{3}}%
\sum_{l=1}^{\infty }l\sin (2\pi \tilde{\beta}l)\left[ \frac{1}{l^{4}}%
+2\sum_{k=1}^{p}\frac{(-1)^{k}\cos (2\pi ka_{0})\cos (\pi k/q)}{%
[l^{2}+4(r/L)^{2}\sin ^{2}(\pi k/q)]^{2}}\right.  \notag \\
&&\left. +\frac{2q}{\pi }\int_{0}^{\infty }dx\frac{h(q,a_{0},2x)\sinh x}{%
\cosh (2qx)-\cos (q\pi )}[l^{2}+4(r/L)^{2}\cosh ^{2}x]^{-2}\right] .
\label{Axialm0}
\end{eqnarray}%
In this case the induced axial current is equal to $(\eta /\alpha )^{4}$
times the corresponding one in Minkowski spacetime in the presence of the
cosmic string (the sign of the axial current obtained in \cite{Mello13} for
the string in Minkowski bulk should be corrected to the opposite one).

Now let us consider the general formula (\ref{Axial3d}) in various
asymptotic regions of the parameters. At large distances from the string, $%
r/\eta \gg 1$, by taking into account that to the leading order $\mathcal{F}%
(q,a_{0},xr^{2}/\eta ^{2})\approx 1$, we see that the current density
coincides with the corresponding result in dS spacetime when the string and
the magnetic flux along the $z$-axis are absent. In the region near the
string, $r/\eta \ll 1$, we use the relation%
\begin{equation}
\mathcal{F}(q,a_{0},z)\approx \frac{q(z/2)^{q(1/2-|a_{0}|)-1/2}}{2\Gamma
(q(1/2-|a_{0}|)+1/2)},  \label{FCalAs}
\end{equation}%
for $z\ll 1$, which directly follows from (\ref{FCal}). To the leading order
this gives%
\begin{eqnarray}
\langle j^{3}\rangle  &=&-\frac{eqL}{\sqrt{2}\pi ^{5/2}\alpha ^{4}}\frac{(r/%
\sqrt{2}\eta )^{q(1-2|a_{0}|)-1}}{\Gamma (q(1/2-|a_{0}|)+1/2)}%
\sum_{l=1}^{\infty }l\sin (2\pi \tilde{\beta}l)  \notag \\
&&\times \int_{0}^{\infty }dx\,x^{q(1/2-|a_{0}|)+1}\,e^{x\left(
1-l^{2}L^{2}/2\eta ^{2}\right) }\mathrm{Re\,}\left[ K_{1/2+im\alpha }(x)%
\right] .  \label{j3Near}
\end{eqnarray}%
The axial current vanishes on the string for $|a_{0}|<(1-1/q)/2$, is finite
for $|a_{0}|=(1-1/q)/2$ and diverges for $|a_{0}|>(1-1/q)/2$. The irregular
mode is responsible for the divergence in the latter case.

For small values of $L/\eta $, the dominant contribution to the axial
current density comes from the part $\langle j^{3}\rangle _{0}$,
corresponding to the current density in the geometry without the cosmic
string and magnetic flux along the $z$-axis, and the contribution of the
string-induced part is exponentially suppressed. By taking into account that
in the limit under consideration the contribution of large $x$ dominates in
the integral of (\ref{j30}), to the leading order we find%
\begin{equation}
\langle j^{3}\rangle \approx -\frac{4\pi ^{-2}e}{\left( \alpha /\eta \right)
^{4}L^{3}}\sum_{l=1}^{\infty }\frac{\sin (2\pi \tilde{\beta}l)}{l^{3}}.
\label{j3smallL}
\end{equation}%
For large values $L/\eta $ with $r/\eta $ fixed, in the integral of (\ref%
{Axial3d}) the contribution from the region near the lower limit dominates.
By making use of (\ref{FCalAs}), for the VEV of the axial current we get%
\begin{eqnarray}
\langle j^{3}\rangle &\approx &-\frac{2eqLB_{2}(m\alpha
)(r/L)^{q(1-2|a_{0}|)-1}}{\pi ^{5/2}(\alpha L/\eta )^{4}\Gamma
(q(1/2-|a_{0}|)+1/2)}  \notag \\
&&\times \sum_{l=1}^{\infty }\frac{\sin (2\pi \tilde{\beta}l)}{%
l^{q(1-2|a_{0}|)+2}}\cos [2m\alpha \ln (lL/\eta )+\beta _{2}(m\alpha )],
\label{j3largeL}
\end{eqnarray}%
where the functions $B_{2}(m\alpha )$ and $\beta _{2}(m\alpha )$ are defined
by the relation (\ref{B2}). As is seen, for the length of the compact
dimension larger than the curvature radius of dS spacetime the axial current
decays as a power-law. For a string in background of Minkowski spacetime and
for a massive fermionic field the axial current density decays as $e^{-mL}$.

\begin{figure}[tbph]
\begin{center}
\begin{tabular}{cc}
\epsfig{figure=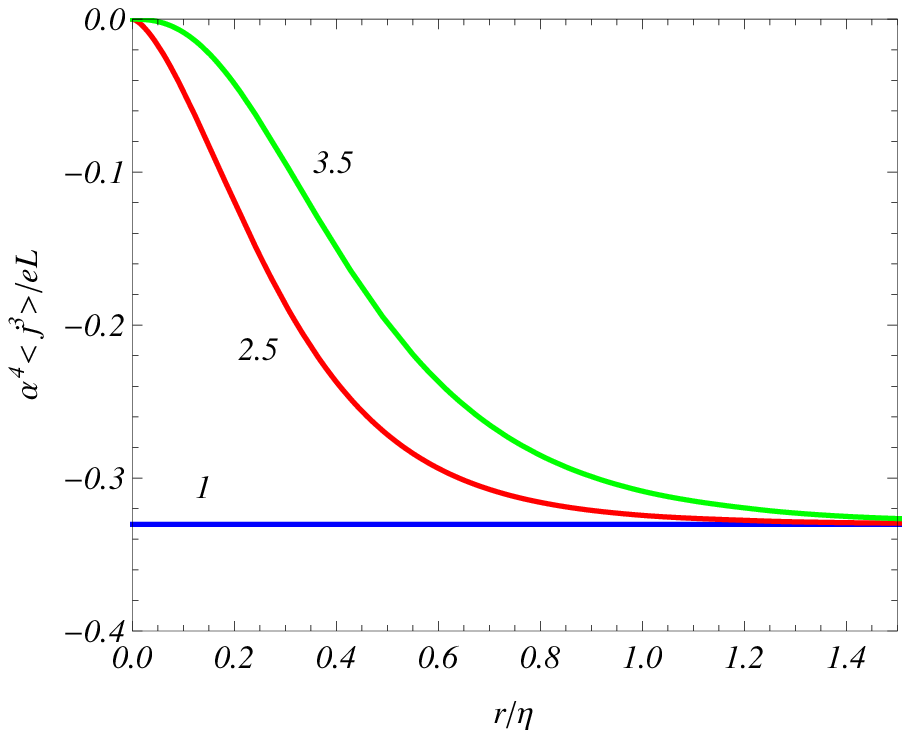,width=7.cm,height=5.5cm} & \quad %
\epsfig{figure=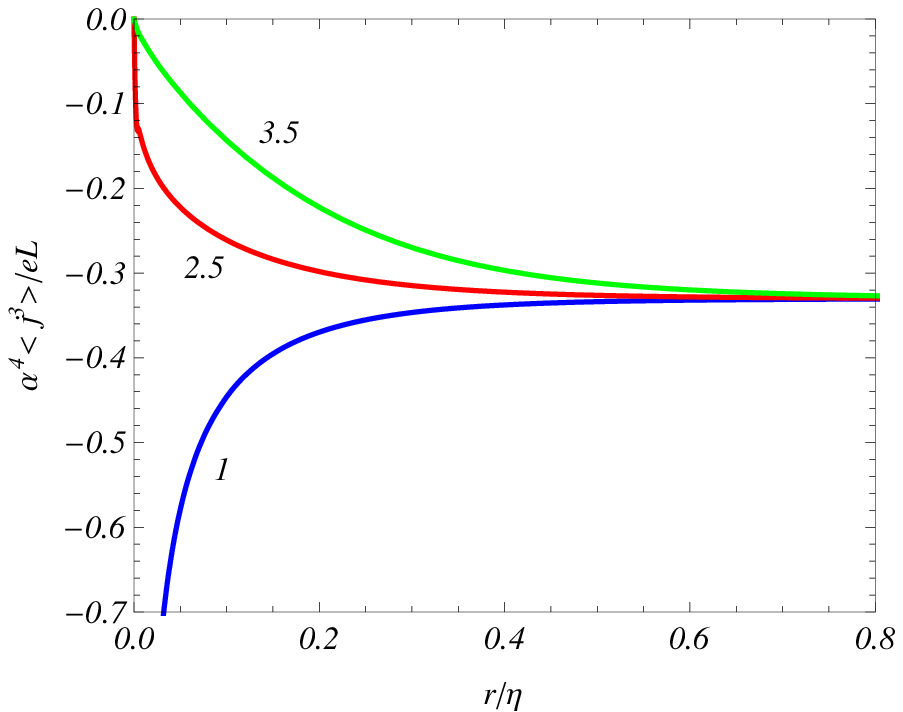,width=7.cm,height=5.5cm}%
\end{tabular}%
\end{center}
\caption{The axial current density as a function of $r/\protect\eta $ for $%
a_{0}=0$ (left panel) and $a_{0}=1/4$ (right panel) and for separate values
of $q$ (numbers near the curves). The graphs are plotted for $m\protect%
\alpha =1$ and $\tilde{\protect\beta}=1/4$.}
\label{fig4}
\end{figure}

In figure \ref{fig4} the axial current density is displayed as a function of
the ratio $r/\eta $ for $a_{0}=0$ (left panel) and $a_{0}=1/4$ (right
panel). The numbers near the curves are the corresponding values of the
parameter $q$. In both panels the graphs are plotted for $m\alpha =1$ and $%
\tilde{\beta}=1/4$. As it has been explained before, at large distances the
axial current density tends to the limiting value corresponding to the
geometry without the cosmic string and in the absence of the magnetic flux
along the $z$-axis (the horizontal line in the left panel corresponding to $%
q=1$ and $a_{0}=0$).

\begin{figure}[tbph]
\begin{center}
\begin{tabular}{cc}
\epsfig{figure=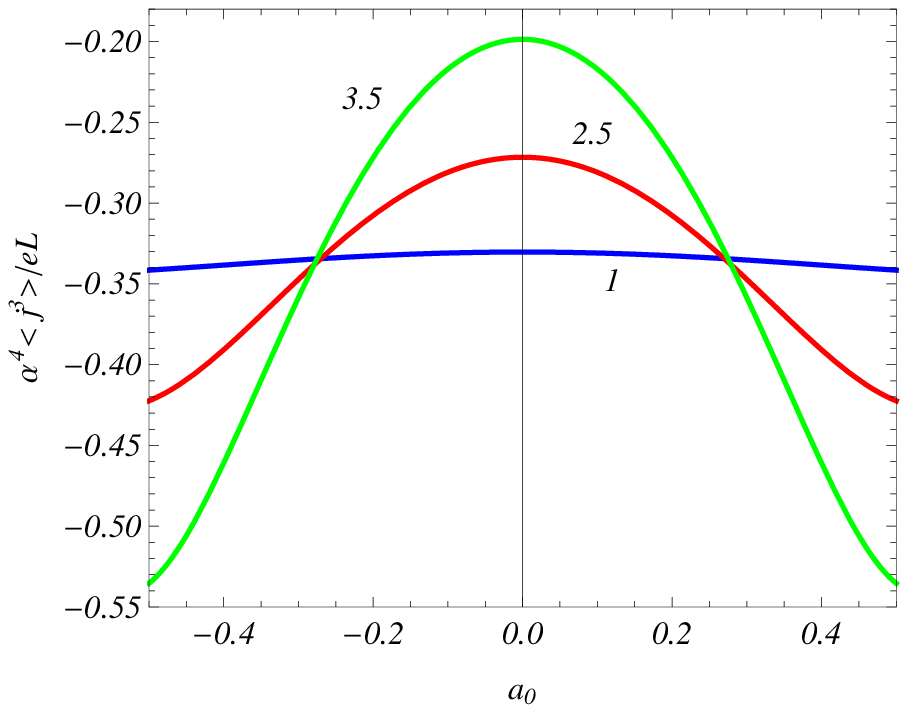,width=7.cm,height=5.5cm} & \quad %
\epsfig{figure=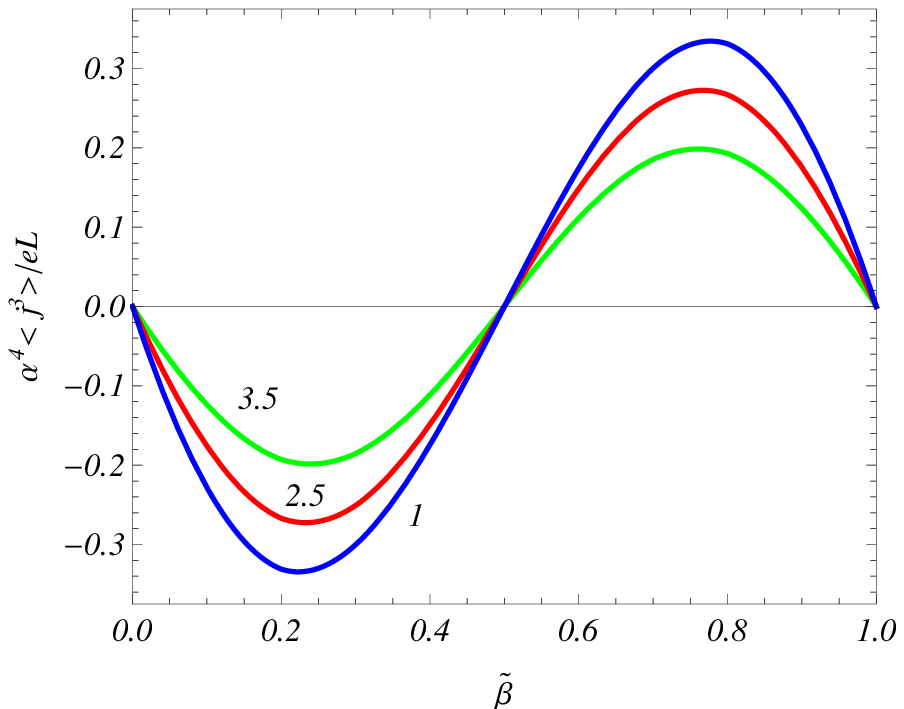,width=7.cm,height=5.5cm}%
\end{tabular}%
\end{center}
\caption{The axial current density versus $a_{0}$ (left panel) and $\tilde{%
\protect\beta}$ (right panel) for different values of the parameter $q$ (the
numbers near the curves). For the left panel $\tilde{\protect\beta}=1/4$ and
for the right one $a_{0}=0$. The graphs are plotted for $L/\protect\eta =1$,
$r/\protect\eta =0.5$, $m\protect\alpha =1$.}
\label{fig5}
\end{figure}

The figure \ref{fig5} presents the dependence of the axial current density
on the parameters $a_{0}$ (left panel) and $\tilde{\beta}$ (right panel) for
separate values of $q$ (numbers near the curves). For the left panel we have
taken $\tilde{\beta}=1/4$ and for the right panel $a_{0}=0$. The values of
the other parameters are as follows: $L/\eta =1$, $r/\eta =0.5$, $m\alpha =1$%
. Note that for fixed values of the other parameters the modulus of the
axial current, $|\langle j^{3}\rangle |$, increases with increasing $q$ for $%
a_{0}$ close to $\pm 1/2$ and decreases for $a_{0}$ close to $0$.

\section{Conclusion}

\label{conc}

In this paper we have investigated the combined effects of the background
gravitational field and topology on the VEV of the current density for a
massive fermionic field. This quantity is an important local characteristic
of the quantum vacuum. In addition to describing the physical structure of a
charged quantum field at a given point, the VEV of the current density acts
as the source in the semiclassical Maxwell equations and plays an important
role in modeling a self-consistent dynamics involving the electromagnetic
field. In order to have an exactly solvable problem, we have taken dS
spacetime as the background geometry. The latter is among the most popular
gravitational backgrounds and plays an important role in cosmology. The
topological effects are induced by a cosmic string and by its
compactification along the axis. Additionally, we have assumed the presence
of a constant gauge field with nonzero axial and azimuthal components. By a
gauge transformation, the problem is reduced to the one in the absence of a
gauge field with the quasiperiodicity conditions (\ref{PerCond1}) and (\ref%
{PerCond2}) on the field operator. In the new representation, the
information about the gauge field is encoded in the phases of these
conditions.

For the evaluation of the VEV of the current density we have employed the
direct summation over the complete set of fermionic modes. For the
Bunch-Davies vacuum state, the corresponding mode functions are given by (%
\ref{psisigma+}). For the model under consideration, in general, there is a
one parameter family of boundary conditions imposed on the field operator at
the location of the string. The fermionic modes used in the present paper
correspond to the boundary condition when the MIT bag boundary condition is
imposed at a finite radius, which is then taken to zero. The formal
expression for the VEV of the induced fermionic current density is presented
in the the mode-sum form (\ref{current}). Because of the compactification of
the cosmic string along its axis, the quantum number corresponding to the $z$%
-direction becomes discrete and for the corresponding summation we use the
Abel-Plana-type formula (\ref{FC3}). As a consequence, the current density
is decomposed in two parts: the first part corresponds to the geometry of a
straight cosmic string in dS spacetime and the second one is induced by the
compactification. For a massless fermionic field, the problem under
consideration is conformally related to the problem with a cosmic string in
Minkowski spacetime and the corresponding expressions for the VEV\ of the
current density are related by the standard conformal transformation.

In the problem under consideration the VEVs of the charge density and the
radial current vanish. For the VEV of the azimuthal current density we have
considered first the part corresponding to the geometry without
compactification, denoted by $\langle j^{2}\rangle _{s}$. Two different
integral representations for this part are provided, the expressions (\ref%
{FC13}) and (\ref{FC9b}). The azimuthal current is an odd periodic function
of the magnetic flux along the string axis with the period equal to the flux
quantum. It depends on the radial and conformal time coordinates through the
ratio $r/\eta $, which is the proper distance from the string measured in
units of the dS curvature scale $\alpha $. Near the string, the leading term
in the asymptotic expansion of $\langle j^{2}\rangle _{s}$ is independent of
the mass and it behaves as the inverse forth power of the proper distance.
In this limit the dominant contribution comes from the modes with small
wavelengths and the effects of the curvature are small. For a massive field,
the influence of the gravitational field on the vacuum current density is
crucial at distances from the string larger than the curvature radius of the
background spacetime. In this limit, corresponding to $r/\eta \gg 1$, the
azimuthal current density exhibits a damping oscillatory behavior with the
amplitude inversely proportional to the fourth power of the distance. Note
that for the string in Minkowski spacetime and for a massive field, at large
distances the current is exponentially suppressed.

The contribution to the azimuthal current density coming from the
compactification of the string along its axis is given by (\ref{FC27}) and
the total current is given by (\ref{j2tot}). In the compactified geometry
the azimuthal current is an odd periodic function of the magnetic flux along
the string and an even periodic function of the flux enclosed by the $z$%
-axis. In both cases the period is equal to the flux quantum. Near the
string the total azimuthal current is dominated by the part $\langle
j^{2}\rangle _{s}$. In this region the leading term in the expansion of
topological part is given by (\ref{j2cnear}). This part vanishes on the
string for $|a_{0}|<(1-1/q)/2$, is finite for $|a_{0}|=(1-1/q)/2$ and
diverges for $|a_{0}|>(1-1/q)/2$. In the opposite limit of large distances
from the string, the behavior of the azimuthal current density for a
compactified cosmic string depends crucially on whether the parameter $%
\tilde{\beta}$, defined by (\ref{abet2}), is zero or not. For $\tilde{\beta}%
\neq 0$ and $q>2$, the leading term in the asymptotic expansion is given by (%
\ref{j2far}) and the azimuthal current is suppressed by the factor $e^{-4\pi
\sigma _{\beta }r\sin (\pi /q)/L}$ with with $\sigma _{\beta }=\min (\tilde{%
\beta},1-\tilde{\beta})$. For $q\leqslant 2$ the suppression is stronger, by
the factor $e^{-4\pi \sigma _{\beta }r/L}$. For $\tilde{\beta}=0$, at large
distances the azimuthal current exhibits a damping oscillatory behavior
described by (\ref{j2far2}). The amplitude of the oscillations decay as $%
(r/\eta )^{-3}$ and in this case the topological part $\langle j^{2}\rangle
_{c}$ dominates in the total current density.

The VEV of the axial current density is given by the expression (\ref%
{Axial3d}). The appearance of the nonzero axial current is a purely
topological effect induced by the compactification of the string along its
axis. The axial current density is an even periodic function of the magnetic
flux along the string axis and an odd periodic function of the flux enclosed
by the $z$-axis with the periods equal to the flux quantum. The modulus of
the axial current, $|\langle j^{3}\rangle |$, increases with increasing $q$
for the values of the parameter $a_{0}$ close to $\pm 1/2$ and decreases for
$a_{0}$ close to $0$. In the absence of the planar angle deficit one has $%
q=1 $ and the general formula is reduced to (\ref{j3q1}). For general case
of the parameter $q$, the corresponding asymptotic near the cosmic string is
given by (\ref{j3Near}) and the axial current density vanishes on the string
for $|a_{0}|<(1-1/q)/2$ and diverges for $|a_{0}|>(1-1/q)/2$. At large
distances from the string the effects of the planar angle deficit and of the
magnetic flux on the axial current are small and to the leading order we
recover the result for dS spacetime with a single compact dimension,
described by (\ref{j30}). For small values of the proper length of the
compact dimension, the axial current is dominated by the part corresponding
to the current density in the geometry without the cosmic string and
magnetic flux along the $z$-axis. In this limit, the leding term in the
asymptotic expansion is given by (\ref{j3smallL}) and the string-induced
corrections to this leading term are exponentially small. For large values
of the ratio $L/\eta $, the behavior of the axial current is described by (%
\ref{j3largeL}). In this range the current density exhibits damping
oscillations with power-law decaying amplitude as a function of the length
of the compact dimension. This behavior is essentially different from the
case of the string in Minkowski bulk with the exponentially suppressed axial
current for large values of $L$.

The results described above can be used, in particular, for the
investigation of the effects induced by cosmic strings in the inflationary
epoch. Though the strings produced before or during early stages of
inflation are diluted by the expansion, cosmic strings can be continuously
formed during inflation by coupling the symmetry breaking and inflaton
fields \cite{Vile94} or by quantum-mechanical tunneling \cite{Basu91}.
Another field of application corresponds to string-driven inflationary
models with the cosmological expansion driven by the string energy \cite%
{Turo88}.

\section*{Acknowledgments}

The authors thank Conselho Nacional de Desenvolvimento Cient\'{\i}fico e
Tecnol\'{o}gico (CNPq) for the financial support. A. A. S. was supported by
the State Committee of Science of the Ministry of Education and Science RA,
within the frame of Grant No. SCS 13-1C040.


\begin{thebibliography}{99}
\bibitem{Birr82} N. D. Birrell and P. C.W. Davies, \textit{Quantum Fields in
Curved Space} (Cambridge University Press, Cambridge, 1982).

\bibitem{Vile94} A. Vilenkin and E.P.S. Shellard, \textit{Cosmic Strings and
Other Topological Defects} (Cambridge University Press, Cambridge, England,
1994).

\bibitem{Damo00} T. Damour and A. Vilenkin, Phys. Rev. Lett. \textbf{85},
3761 (2000); P. Battacharjee and G. Sigl, Phys. Rep. \textbf{327}, 109
(2000); V. Berezinski, B. Hnatyk, and A. Vilenkin, Phys. Rev. D \textbf{64},
043004 (2001).

\bibitem{Sara02} S. Sarangi and S.H.H. Tye, Phys. Lett. B \textbf{536}, 185
(2002); E.J. Copeland, R.C. Myers, and J. Polchinski, JHEP \textbf{06}, 013
(2004); G. Dvali and A. Vilenkin, JCAP \textbf{0403}, 010 (2004); J.
Polchinski, arXiv:hep-th/0410082.

\bibitem{Hell86} T.M. Helliwell and D.A. Konkowski, Phys. Rev. D \textbf{34}%
, 1918 (1986).

\bibitem{Hisc87} W.A. Hiscock, Phys. Lett. B \textbf{188}, 317 (1987).

\bibitem{Line87} B. Linet, Phys. Rev. D \textbf{35}, 536 (1987).

\bibitem{Frol87} V.P. Frolov and E.M. Serebriany, Phys. Rev. D \textbf{35},
3779 (1987).

\bibitem{Dowk87} J.S. Dowker, Phys. Rev. D \textbf{36}, 3095 (1987); J.S.
Dowker, Phys. Rev. D \textbf{36}, 3742 (1987).

\bibitem{Smit89} A.G. Smith, in \textit{The Formation and Evolution of
Cosmic Strings}, Proceedings of the Cambridge Workshop, Cambridge, England,
1989, edited by G.W. Gibbons, S.W. Hawking, and T. Vachaspati (Cambridge
University Press, Cambridge, England, 1990).

\bibitem{Mats90} G.E.A. Matsas, Phys. Rev. D \textbf{41}, 3846 (1990).

\bibitem{Alle90} B. Allen and A.C. Ottewill, Phys. Rev. D \textbf{42}, 2669
(1990); B. Allen, J.G. Mc Laughlin, and A.C. Ottewill, Phys. Rev. D \textbf{%
45}, 4486 (1992).

\bibitem{Sour92} T. Souradeep and V. Sahni, Phys. Rev. D \textbf{46}, 1616
(1992).

\bibitem{Shir92} K. Shiraishi and S. Hirenzaki, Class. Quantum Grav. \textbf{%
9}, 2277 (1992).

\bibitem{Beze94} V.B. Bezerra and E.R. Bezerra de Mello, Class. Quantum
Grav. \textbf{11}, 457 (1994); E.R. Bezerra de Mello, Class. Quantum Grav.
\textbf{11}, 1415 (1994).

\bibitem{Furs94} D. Fursaev, Class. Quantum Grav. \textbf{11}, 1431 (1994).

\bibitem{Cogn94} G. Cognola, K. Kirsten, and L. Vanzo, Phys. Rev. D \textbf{%
49}, 1029 (1994).

\bibitem{Guim94} M.E.X. Guimar\~{a}es and B. Linet, Commun. Math. Phys.
\textbf{165}, 297 (1994).

\bibitem{Line95} B. Linet, J. Math. Phys. \textbf{36}, 3694 (1995).

\bibitem{More95} E.S. Moreira Jnr, Nucl. Phys. B \textbf{451}, 365 (1995).

\bibitem{Alle96} B. Allen, B.S. Kay, and A.C. Ottewill, Phys. Rev. D \textbf{%
53}, 6829 (1996).

\bibitem{Bord96a} M. Bordag, K. Kirsten, and S. Dowker, Commun. Math. Phys.
\textbf{182}, 371 (1996).

\bibitem{Iell97} D. Iellici, Class. Quantum Grav. \textbf{14}, 3287 (1997).

\bibitem{Khus99} N.R. Khusnutdinov and M. Bordag, Phys. Rev. D \textbf{59},
064017 (1999).

\bibitem{Spin03} J. Spinelly and E.R. Bezerra de Mello, Class. Quantum Grav.
\textbf{20}, 873 (2003).

\bibitem{Beze06} V.B. Bezerra and N.R. Khusnutdinov, Class. Quantum Grav.
\textbf{23}, 3449 (2006).

\bibitem{Beze06b} E.R. Bezerra de Mello, V.B. Bezerra, A.A. Saharian, and
A.S. Tarloyan, Phys. Rev. D \textbf{74}, 025017 (2006); E.R. Bezerra de
Mello, V.B. Bezerraa, and A.A. Saharian, Phys. Lett. B \textbf{645}, 245
(2007); E.R. Bezerra de Mello, V.B. Bezerra, A.A. Saharian, and A.S.
Tarloyan, Phys. Rev. D \textbf{78}, 105007 (2008).

\bibitem{charged1} M.E.X. Guimar\~{a}es and B. Linet, Commun. Math. Phys.
\textbf{165}, 297 (1994).

\bibitem{charged3} J. Spinelly and E.R. Bezerra de Mello, Class. Quantum
Grav. \textbf{20} 874, (2003); J. Spinelly and E.R. Bezerra de Mello, Int.
J. Mod. Phys. A, \textbf{17}, 4375 (2002).

\bibitem{Spin} J. Spinelly and E.R. Bezerra de Mello, Int. J. Mod. Phys. D
\textbf{13}, 607 (2004); J. Spinelly and E.R. Bezerra de Mello, Nucl Phys. B
(Proc. Suppl.) \textbf{127}, 77 (2004).

\bibitem{Spin1} J. Spinelly and E. R. Bezerra de Mello, JHEP \textbf{09},
005 (2008).

\bibitem{Site12} Yu.A. Sitenko and N.D. Vlasii, Class. Quantum Grav. \textbf{%
29}, 095002 (2012).

\bibitem{Sira} L. Sriramkumar, Class. Quantum Grav. \textbf{18}, 1015 (2001).

\bibitem{Yu} Yu.A. Sitenko and N.D. Vlasii, Class. Quantum Grav. \textbf{26}%
, 195009 (2009).

\bibitem{Mello10} E.R. Bezerra de Mello, Class. Quantum Grav. \textbf{27},
095017 (2010).

\bibitem{Saha10} E.R. Bezerra de Mello, V.B. Bezerra, A.A. Saharian, and
V.M. Bardeghyan, Phys. Rev. D \textbf{82}, 085033 (2010).

\bibitem{Davi88} P.C.W. Davies and V. Sahni, Class. Quantum Grav. \textbf{5}%
, 1 (1988).

\bibitem{Otte10} A.C. Ottewill and P. Taylor, Phys. Rev. D \textbf{82},
104013 (2010); A.C. Ottewill and P. Taylor, Class. Quantum Grav. \textbf{28}%
, 015007 (2011).

\bibitem{Beze09} E.R. Bezerra de Mello and A.A. Saharian, JHEP \textbf{04},
046 (2009).

\bibitem{Beze10} E.R. Bezerra de Mello and A.A. Saharian, JHEP \textbf{08},
038 (2010).

\bibitem{Beze12} E.R. Bezerra de Mello and A.A. Saharian, J. Phys. A: Math.
Theor. \textbf{45} 115402 (2012).

\bibitem{Beze13} E.R. Bezerra de Mello and A.A. Saharian, Class. Quantum.
Grav. \textbf{30} 175001 (2013).

\bibitem{Cast09} A.H. Castro Neto, F. Guinea, N.M.R. Peres, K.S. Novoselov,
and A.K. Geim, Rev. Mod. Phys. \textbf{81}, 109 (2009).

\bibitem{Mello13} E. R. Bezerra de Mello and A. A. Saharian, Eur. Phys. J.
C. \textbf{73}, 2532 (2013).

\bibitem{Bellu14} S. Bellucci, E. R. Bezerra de Mello, A. de Padua, and A.
A. Saharian, Eur. Phys. J. C. \textbf{74}, 2688 (2014).

\bibitem{Bell10FC} S. Bellucci, A.A. Saharian, and V.M. Bardeghyan, Phys.
Rev. D \textbf{82}, 065011 (2010); S. Bellucci and A.A. Saharian, Phys. Rev.
D \textbf{87}, 025005 (2013).

\bibitem{Bell12} S. Bellucci, A.A. Saharian, and H.A. Nersisyan, Phys. Rev.
D \textbf{88}, 024028 (2013).

\bibitem{Beze13SC} E.R. Bezerra de Mello and A.A. Saharian, Phys. Rev. D
\textbf{87}, 045015 (2013)

\bibitem{Bell14b} S. Bellucci, E.R. Bezerra de Mello, and A.A. Saharian,
Phys. Rev. D \textbf{89}, 085002 (2014).

\bibitem{Ghez02} A. M. Ghezelbash and R.B. Mann, Phys. Lett. B \textbf{537},
329 (2002).

\bibitem{B-D} J.D. Bjorken and S.D. Drell, \textit{Relativistic Quantum
Mechanics} (McGraw-Hill, New York, 1964).

\bibitem{Bunc78} T.S. Bunch and P.C.W. Davies, Proc. R. Soc. London A
\textbf{360}, 117 (1978).

\bibitem{Sous89} P. de Sousa Gerbert and R. Jackiw, Commun. Math. Phys.
\textbf{124}, 229 (1989); P. de Sousa Gerbert, Phys. Rev. D \textbf{40},
1346 (1989); Yu.A. Sitenko, Ann. Phys. \textbf{282}, 167 (2000).

\bibitem{Bene00} C.G. Beneventano, M. De Francia, K. Kirsten, and E.M.
Santangelo, Phys. Rev. D \textbf{61}, 085019 (2000); M. De Francia and K.
Kirsten, Phys. Rev. D \textbf{64}, 065021 (2001).

\bibitem{Bell11} S. Bellucci, E.R. Bezerra de Mello, and A.A. Saharian,
Phys. Rev. D \textbf{83}, 085017 (2011); E.R. Bezerra de Mello, F. Moraes,
A.A. Saharian, Phys. Rev. D \textbf{85}, 045016 (2012).

\bibitem{Site99} Yu.A. Sitenko, Phys. Rev. D \textbf{60}, 125017 (1999).

\bibitem{Abra64} M. Abramowitz and I.A. Stegun, \textit{Handbook of
Mathematical Functions} (National Bureau of Standards, Washington, DC, 1972).

\bibitem{SahaRev} A.A. Saharian, \textit{The Generalized Abel-Plana Formula
with Applications to Bessel Functions and Casimir Effect} (Yerevan State
University Publishing House, Yerevan, 2008); Preprint ICTP/2007/082;
arXiv:0708.1187.

\bibitem{Beze08} E.R. Bezerra de Mello and A.A. Saharian, Phys. Rev. D
\textbf{78}, 045021 (2008); S. Bellucci and A.A. Saharian, Phys. Rev. D
\textbf{79}, 085019 (2009).

\bibitem{Wats44} G.N. Watson, \textit{A Treatise on the Theory of Bessel
Functions} (Cambridge, Cambridge University Press, 1944).

\bibitem{Gradshteyn} I.S. Gradshteyn and I.M. Ryzhik, \textit{Table of
Integrals, Series and Products} (Academic Press, New York, 1980).

\bibitem{Basu91} R. Basu, A.H. Guth, and A. Vilenkin, Phys. Rev. D \textbf{44%
}, 340 (1991).

\bibitem{Turo88} N. Turok, Phys. Rev. Lett. \textbf{60}, 549 (1988).
\end{thebibliography}
\end{document}